%% file: 0_main.tex
\documentclass[sigplan,twocolumn,nonacm]{acmart}
\acmSubmissionID{<610>}
\renewcommand\footnotetextcopyrightpermission[1]{}

\newcommand{\name}{\textsc{ElasWave}}
\newcommand{\namerc}{\textsc{ReCycle}}
\newcommand{\nametft}{\textsc{TorchFT}}
\newcommand{\stitle}[1]{\vspace{1ex} \noindent{{\bf #1}}}
\usepackage{lineno}
\usepackage{titlesec}
\usepackage{subcaption}
\usepackage{graphicx}
\usepackage{amsmath,amsfonts}
\usepackage{algorithm}
\usepackage{textcomp}
\usepackage{xcolor}
\usepackage{pifont}
\usepackage{hyperref}
\usepackage{algpseudocode}
\usepackage{float}     
\usepackage{placeins}  
\usepackage{mathtools}
\usepackage{booktabs}
\usepackage{multirow}    
\newcommand{\cmark}{$\checkmark$} 
\newcommand{\maybe}{$\triangle$}  
\newcommand{\xmark}{$\times$}     
\usepackage{longtable}
\usepackage{mweights} 

\titleformat{\subsubsection}[runin]
  {\normalfont\bfseries}{\thesubsubsection}{1em}{}
\AtBeginDocument{%
  }

\title{\textsc{ElasWave}: An Elastic-Native System for Scalable Hybrid-Parallel Training}

\begin{document}

\author{\vspace{-0.6em}%
  \large
  Xueze Kang$^{1\dagger}$,\enspace
  Guangyu Xiang$^{1\dagger}$, \enspace
  Yuxin Wang$^{4*}$, \enspace
  Hao Zhang$^{4}$,\enspace
  Yuchu Fang$^{4}$,\enspace 
  Yuhang Zhou$^{3}$,\enspace \\
  Zhenheng Tang$^{2}$,\enspace
  Youhui Lv$^{5}$,\enspace
  Eliran Maman$^{7}$,\enspace
  Mark Wasserman$^{7}$,\enspace
  Alon Zameret$^{7}$,\enspace 
  Zhipeng Bian$^{6}$,\enspace \\
  Shushu Chen$^{5}$,\enspace 
  Zhiyou Yu$^{5}$,\enspace
  Jin Wang$^{5}$,\enspace
  Xiaoyu Wu$^{4}$,  \enspace
  Yang Zheng$^{4}$,  \enspace
  Chen Tian$^{3}$,  \enspace
  Xiaowen Chu$^{1*}$\\[2pt]
  \normalsize $^{1}$\textit{HKUST(GZ)} \quad
  $^{2}$\textit{HKUST} \quad
  $^{3}$\textit{NJU}\\
  \textit{Huawei} \enspace
  $^{4}$\textit{TTE Lab} \quad
  $^{5}$\textit{ICT BG} \quad
  $^{6}$\textit{Cloud} \quad
  $^{7}$\textit{TRC Team} 
}

\input{Contents/0_abstract}


\maketitle
\renewcommand{\shortauthors}{}

\begingroup
\renewcommand\thefootnote{\fnsymbol{footnote}}
\footnotetext[2]{Equal contribution.}
\footnotetext[1]{Corresponding authors: \texttt{yuxin.wang11@huawei.com}, \texttt{xwchu@hkust-gz.\\edu.cn}.}
\endgroup

\input{Contents/1_Introduction}

\input{Contents/2_Background}

\input{Contents/3_Overview}

\input{Contents/5_EGS}

\input{Contents/4_ZCE}
\input{Contents/6_CRR}

\input{Contents/8_Evaluation}

\input{Contents/9_Related_Work}
\input{Contents/10_Conclusion}
\newpage
\bibliographystyle{ACM-Reference-Format}
\bibliography{sample-base}


\end{document}

%% file: Contents/0_abstract.tex
 \begin{abstract}
Large-scale LLM pretraining now runs across $10^5$–$10^6$ accelerators, making failures routine and elasticity mandatory. We posit that an elastic-native training system must jointly deliver (i) parameter consistency, (ii) low mean time to recovery (MTTR), (iii) high post-change throughput, and (iv) computation consistency. No prior system achieves all four simultaneously.
To achieve these goals, we present \name, which delivers per-step fault tolerance via multi-dimensional scheduling across graph, dataflow, DVFS, and RNG. \name~reshapes and reshards micro-batches while preserving the global batch size and gradient scale. It performs online pipeline resharding with asynchronous parameter migration and interleaves ZeRO partitions, reducing parameter recovery processes to disjoint rank-to-rank transfers. 
It further leverages DVFS to absorb pipeline bubbles and reshards RNG to keep computation consistency.
Together, a dynamic communicator enables in-place communication group edits, while per-step in-memory snapshots support online verification and redistribution. We evaluate \name~on 96 NPUs and benchmark it against state-of-the-art baselines: throughput improves by $1.35\times$ over \namerc~and $1.60\times$ over \nametft; communicator recovery completes within one second (up to $82\times/3.6\times$ faster than full/partial rebuilds); migration MTTR drops by as much as $51\%$; and convergence deviation is reduced by approximately $78\%$.


\end{abstract}

%% file: Contents/1_Introduction.tex
\section{Introduction}
\begin{table*}[t]
  \centering
  \caption{Capability matrix of elastic-training systems. Columns summarize support for: \textit{MTTR}, \textit{Throughput}, \textit{Computation Consistency}, and \textit{Parameter Consistency}. Symbols: \cmark supported, × unsupported, \maybe partial. \name~is the only system covering DP\&PP granularity with joint data/graph/hardware planning, RNG consistency, and both rollback and live resharding.}
  \resizebox{2\columnwidth}{!}{%
  \begin{tabular}{|c|c|c|c|c|c|c|c|c|c|c|}
    \toprule
          & \multicolumn{2}{c|}{\textbf{MTTR}} & \multicolumn{4}{c|}{\textbf{Throughput}} & \multicolumn{2}{c|}{\textbf{Computation Consistency}} & \multicolumn{2}{c|}{\textbf{Parameter Consistency}} \\
    \midrule
    \textbf{System} & \textbf{Online} & \textbf{Optim.} & \textbf{Granularity} & \textbf{Data} & \textbf{Graph} & \textbf{Hardware} & \textbf{Systematic} & \textbf{Numerical} & \textbf{Per-step Rollback} & \textbf{Live Reshard} \\
    \midrule
    TorchElastic~\cite{torchelastic} & \multirow{4}[8]{*}{\xmark} & \multirow{5}[10]{*}{\xmark} & DP    & \multirow{5}[10]{*}{\xmark} & \multirow{4}[8]{*}{\xmark} & \multirow{7}[14]{*}{\xmark} & \multirow{3}[6]{*}{\xmark} & \multirow{7}[14]{*}{\xmark} & \xmark & \multirow{7}[14]{*}{\xmark} \\
\cmidrule{1-1}\cmidrule{4-4}\cmidrule{10-10}    DLRover~\cite{wang2024dlrover} &       &       & DP    &       &       &       &       &       & \maybe &  \\
\cmidrule{1-1}\cmidrule{4-4}\cmidrule{10-10}    ByteCheckpoint~\cite{byteckpt_nsdi25} &       &       & \xmark &       &       &       &       &       & \cmark &  \\
\cmidrule{1-1}\cmidrule{4-4}\cmidrule{8-8}\cmidrule{10-10}    EasyScale~\cite{li2023easy} &       &       & DP    &       &       &       & \cmark &       & \multirow{4}[8]{*}{\xmark} &  \\
\cmidrule{1-2}\cmidrule{4-4}\cmidrule{6-6}\cmidrule{8-8}    Oobleck~\cite{oobleck_sosp23} & \cmark &       & DP\&PP &       & \cmark &       & \xmark &       &       &  \\
\cmidrule{1-6}\cmidrule{8-8}    \namerc~\cite{recycle_sosp24} & \cmark & \maybe & DP\&PP & \cmark & \multirow{2}[4]{*}{\xmark} &       & \maybe &       &       &  \\
\cmidrule{1-5}\cmidrule{8-8}    \nametft~\cite{torchft_blog} & \cmark & \xmark & DP    & \xmark &       &       & \xmark &       &       &  \\
    \midrule
    \textbf{\name} & \textbf{\cmark} & \textbf{\cmark} & \textbf{DP\&PP} & \textbf{\cmark} & \textbf{\cmark} & \textbf{\cmark} & \textbf{\maybe} & \textbf{\cmark} & \textbf{\cmark} & \textbf{\cmark} \\
    \bottomrule
  \end{tabular}%
  }
  \label{tab:allmethod}
\end{table*}

LLM training ~\cite{xai_grok4_2025,meta_llama4_blog_2025,team2024qwen2,comanici2025gemini,liu2024deepseek} has progressed from tens of thousands to $10^5$ accelerators, with roadmaps aiming at $10^6$ (e.g., the Stargate system)~\cite{openai_stargate_uae_2025}. At current scale, Google Cloud grows a single 8{,}960-chip TPU pod to tens of thousands via Multislice~\cite{gcloud_tpu_v6e_zh_2025}, while Huawei’s UnifiedBus-based Atlas SuperPoDs (8{,}192/15{,}488 NPUs) aggregate into SuperClusters at $5\times10^5$ to $10^6$ scales~\cite{huawei_lingqu_superpod_2025}.
With hyperscale clusters and preemptible clouds, month-long pretraining is routine where fail-stop~\cite{kokolis2025,byteckpt_nsdi25,jiang2024megascale,weng2022mlaas, hu2024characterization} and fail-slow~\cite{wu2025adaptra, wu2024falcon,lin2025understanding} events are commonplace. For instance, a 16{,}384-H100 Llama-3 run reported interruptions roughly every three hours, with nearly half due to GPU or HBM issues~\cite{kokolis2025}. Elastic training is therefore a first-class requirement: Systems must maintain progress despite resource variation, reconfigure~\cite{oobleck_sosp23,recycle_sosp24,thorpe2023bamboo}, sustain throughput~\cite{recycle_sosp24}, and preserve convergence~\cite{pollux_osdi21,li2023easy}.

We define four objectives for an elastic-native training system in production: (i) \textit{Parameter Consistency} across hybrid parallelism, (ii) low \textit{Mean Time To Recovery (MTTR)}, (iii) high \textit{Throughput} after scale-in/scale-out, and (iv) \textit{Computation Consistency} that preserves the optimization trajectory of a static run~\cite{li2023easy}. These objectives are coupled, and no prior system achieves all four simultaneously. \textsc{Oobleck}~\cite{oobleck_sosp23} ensures parameter consistency via redundancy but imposes steady-state overhead. Other works~\cite{li2023easy}\cite{recycle_sosp24} preserve DP determinism, whereas do not guarantee consistency for sharded state\allowbreak~\cite{byteckpt_nsdi25}. 
The academic state-of-the-art (SOTA), \textsc{ReCycle}~\cite{recycle_sosp24} avoids restart by using pipeline bubbles for intra-stage rerouting. 
However, when training scales, the tiny bubble budget is quickly exhausted, leading to stragglers and out-of-memory (OOMs). 
The industrial SOTA, \textsc{TorchFT}~\cite{torchft_blog}, sustains step-level (i.e., one training iteration) progress by dropping and rejoining DP replicas, which wastes substantial compute resources and creates pronounced throughput cliffs~\cite{torchft_blog}.

This paper presents \name~as a comprehensive solution. It introduces \emph{multi-dimensional scheduling} that delivers per-step fault tolerance by coordinating four axes—dataflow, graph, accelelator frequency, and RNG. 
(i) \textit{Dataflow}: we resize and reshard micro batches while preserving global batch size and gradient scale for immediate continuation. 
(ii) \textit{Graph}: we propose online pipeline resharding with layer migration to restore load balance. We introduce non-blocking migration to overlap with compute, and interleave ZeRO state sharding to reduce recovery to disjoint rank-to-rank sends, avoiding intra-group reshaping and large transfers. 
(iii) \textit{Hardware}: frequency scaling removes residual bubbles after parameter recovery. 
(iv) \textit{RNG}: RNG resharding maintains step-identical random streams after membership changes and bounds numerical drift. 
To further improve recovery efficiency, we also introduce a dynamic communicator that reuses existing connections to avoid full rebuilds. 
Together with per-step in-memory snapshots and live remapping, parameters are verified, redistributed, and loaded on-the-fly to achieve efficient elasticity. As shown in Table~\ref{tab:allmethod}, \name~is, to our knowledge, the first scalable, elastic-native training system on XPU clusters.

\name~is built with two components: \textit{\name~Agent} and \textit{\name~Core}. The Agent detects interruptions ; the Core plans and executes elastic responses:

\textbf{(i) Scheduling:} \name~Core performs elastic multi-dimensional scheduling within a single step, jointly deciding \emph{Dataflow, Graph, DVFS, RNG} and emitting an \emph{executable recovery plan} that optimizes parameter consistency, low MTTR, post-change throughput, and computation consistency.

 \textbf{(ii) Executions:} \name~Core minimizes MTTR by (i) maintaining step-level in-memory snapshots with live remap, (ii) using a \emph{Dynamic Communicator} to edit groups in place for sub-second recovery, and (iii) \emph{overlapping} interleaved stage resharding to avoid blocking training.
 
\textbf{(i) Implementations:} \name~provides end-to-end deployment on an Ascend-910B NPU cluster, with hierarchical interfaces designed on CANN~\cite{huawei_cann_intro_2025_misc}, Torch-npu~\cite{paszke2019pytorch}, Megatron~\cite{narayanan2021} with 20k+ loc.

 \textbf{(iv) Results:} We evaluate \name~in LLM training with SOTA hybrid parallel setups~\cite{10.1145/3620666.3651359,rajbhandari2020zero} across production traces. On our testbed, the throughput gains outperform \textsc{ReCycle}~\cite{recycle_sosp24} by $1.35\times$ and \textsc{TorchFT}~\cite{torchft_blog} by $1.60\times$, while preserving parameter consistency. For MTTR, communicator recovery completes in sub-second time, improved by up to $82\times$ and $3.6\times$ compared with full and partial rebuilds method. Non-blocking migration with interleaved ZeRO cuts layer migration MTTR by up to 51\% compared with blocked migration under vanilla ZeRO. RNG Resharding reduces convergence deviation by ~78\%, improving computation consistency.

%% file: Contents/2_Background.tex
\section{Principles of An Elastic-Native System}\label{sec:multi-dimensional}
\begin{figure}[b]
  \centering
  \includegraphics[width=1\columnwidth]{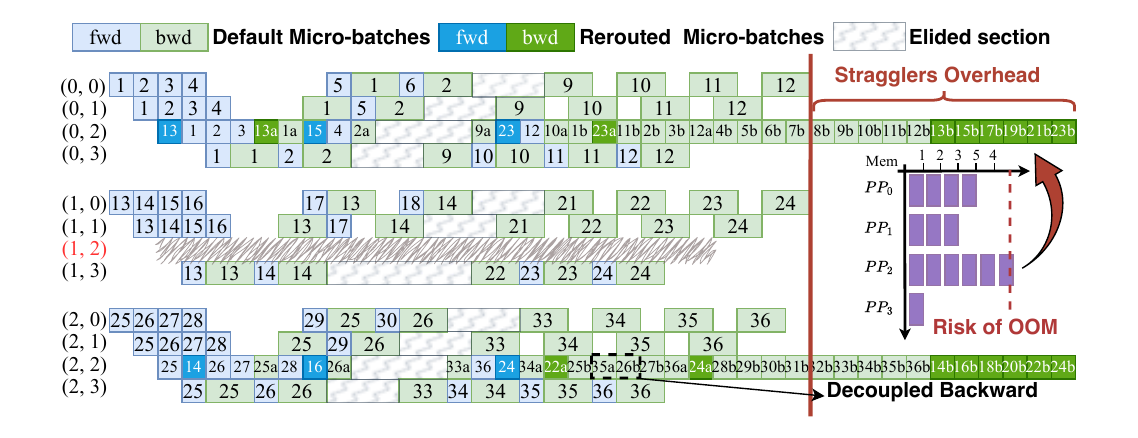}
  \caption{Pipeline schedule for \namerc~after a failure at node (1,2). \namerc~reroutes the failed rank's work to peers in the same stage (e.g., (0,2) and (2,2)), creating a straggler. While its decoupled backward pass creates bubbles to absorb the extra work, the large number of rerouted micro-batches quickly exhausts this bubble budget. The strategy also extends activation lifetimes, which increases memory pressure and risks Out-of-Memory (OOM) errors.}
  \label{fig:recycle}
\end{figure}

\begin{figure*}[!t]
    \centering
    \includegraphics[width=1\textwidth]{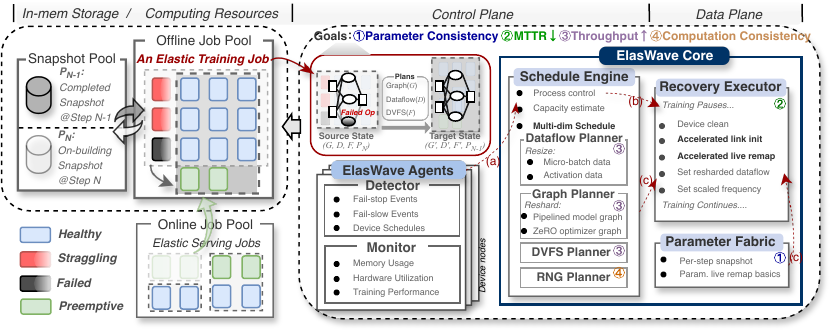}
    \caption{System architecture of \name, illustrating the elastic recovery workflow and its triggers (fail-stop, fail-slow, and resource-scheduling signals). (a) When the \name~Agent detects a failure, straggler, or scheduling signals, it reports the current cluster state to the \name~Core. The Core then generates a multi-dim plan to optimize four key goals. (b) The Engine first pauses the training job via the Recovery Executor. (c) The recovery plan is dispatched to the Recovery Executor. The Executor uses the plan to perform an accelerated live remap, reconfigure links, and set the new dataflow, using state provided by the Parameter Fabric from the in-memory Snapshot Pool. Once the cluster is reconfigured, training resumes.}
    \label{fig:System_Arch}
\end{figure*}

\begin{figure}
    \centering
    \includegraphics[width=\linewidth]{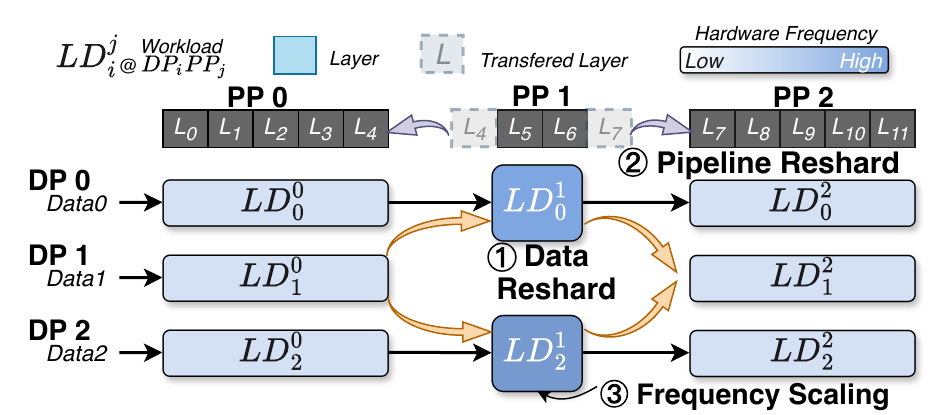}
      \caption{To optimize throughput, \name's multi-dim scheduling combines three strategies. After a failure, it first performs \textit{Data Reshard} in DP domain (\textcircled{1}), then uses \textit{Pipeline Reshard} in PP domain (\textcircled{2}) to balance the workload, and finally eliminates remaining pipeline bubbles by \textit{DVFS} (\textcircled{3}).}
    \label{fig:rebalance}
\end{figure}

Building an elastic-native system requires end-to-end consideration during the system design. The system must simultaneously deliver per-step parameter consistency, low MTTR, high throughput, and convergence consistency. Figure~\ref{fig:recycle} is an example in \namerc~\cite{recycle_sosp24} that achieves fast failure recovery but introduces OOM and straggler issues with cummulative micro-batches in the cool-down phase in production.

\textbf{Efficient Recovery with Consistent Parameters.} Reconfiguring hybrid-parallel training (DP+PP+ZeRO/FSDP) \emph{on the fly}, without a restart, is challenging: it requires reshaping PP by redistributing layers, splitting/merging DP groups to adjust replication, and reassigning sharded optimizer/gradient states. These operations are typically heavy-weight, the recovery process of which includes large parameter migration and global coordination to overlap communication with compute. Prior systems either freeze the logical layout~\cite{li2023easy}, precompute a limited set of templates~\cite{oobleck_sosp23}, or handle only intra-stage DP failover given a surviving replica~\cite{recycle_sosp24}, leaving production training without end-to-end sharding/loading solutions.

\textit{Takeaway.} An elastic-native system must generalize to arbitrary scale-in/out via precise state management and accelerated transfers \emph{without} interrupting the next update. Recovery time should be itemized by component and minimized.

\textbf{Computation Consistency.} Elastic scaling can silently perturb both statistical and numerical consistency relative to a fixed-configuration run. To preserve trust, the effective batch size, RNG streams, execution order, and communication/reduction ordering must remain minimally perturbed; otherwise elasticity risks loss spikes or silent data corruptions~\cite{li2023easy}. We therefore treat \emph{convergence invariance} as the first constraint in our elastic scheduler.

\textit{Takeaway.} In \name, we keep the computation graph and dataflow as regular as possible so elasticity remains traceable, and we implement deterministic hybrid-parallel RNG remapping aligned with each scheduling decision.

\textbf{Maximize Throughput Under Constraints.} Achieving high post-change throughput is a multi-objective scheduling problem: decisions in one dimension can conflict with others. Meeting all four goals therefore requires a careful systems design that coordinates these dimensions and integrates ideas from fault tolerance (fast recovery~\cite{wang2023gemini,gupta2024just}, redundancy~\cite{thorpe2023bamboo}), high-performance computing (load balancing~\cite{zhang2024hap, qi2024pipeline}, scheduling~\cite{liu2024aceso,zheng2022alpa}), and ML engineering (statistical validity~\cite{pollux_osdi21}, reproducibility~\cite{semmelrock2025reproducibility}) into a single framework.

\textit{Takeaway.} Guided by multi-dimensional scheduling, our system jointly reasons about data, model, hardware, and RNG to navigate trade-offs that one-dimensional approaches cannot handle. \name~introduces online planners that, upon each resource change, compute a new hybrid-parallel plan (including device mappings) to maximize throughput under the current hardware pool, and seamlessly transition the running job to that plan. 

The following sections will detail how our proposed system addresses each of these challenges, enabling on-the-fly elastic training for LLMs without sacrificing availability, efficiency, or model quality.

%% file: Contents/3_Overview.tex
\section{System Overview}
\name~is an elastic-native LLM training system that delivers per-step fault tolerance via multi-dimensional scheduling across four axes: data, computation graph, device frequency, and RNG (Figure \ref{fig:System_Arch}). It delivers per-step fault tolerance while jointly delivering (1)\textit{parameter consistency} (across DP/PP/TP with ZeRO/FSDP states), (2)\textit{low mean time to recovery (MTTR)} at disturbances, (3)\textit{high post-change throughput}, and (4)\textit{computation consistency}—the resumed run follows the same optimization trajectory as fault-free training.
\begin{figure*}[!t]
    \centering
    \includegraphics[width=1\linewidth]{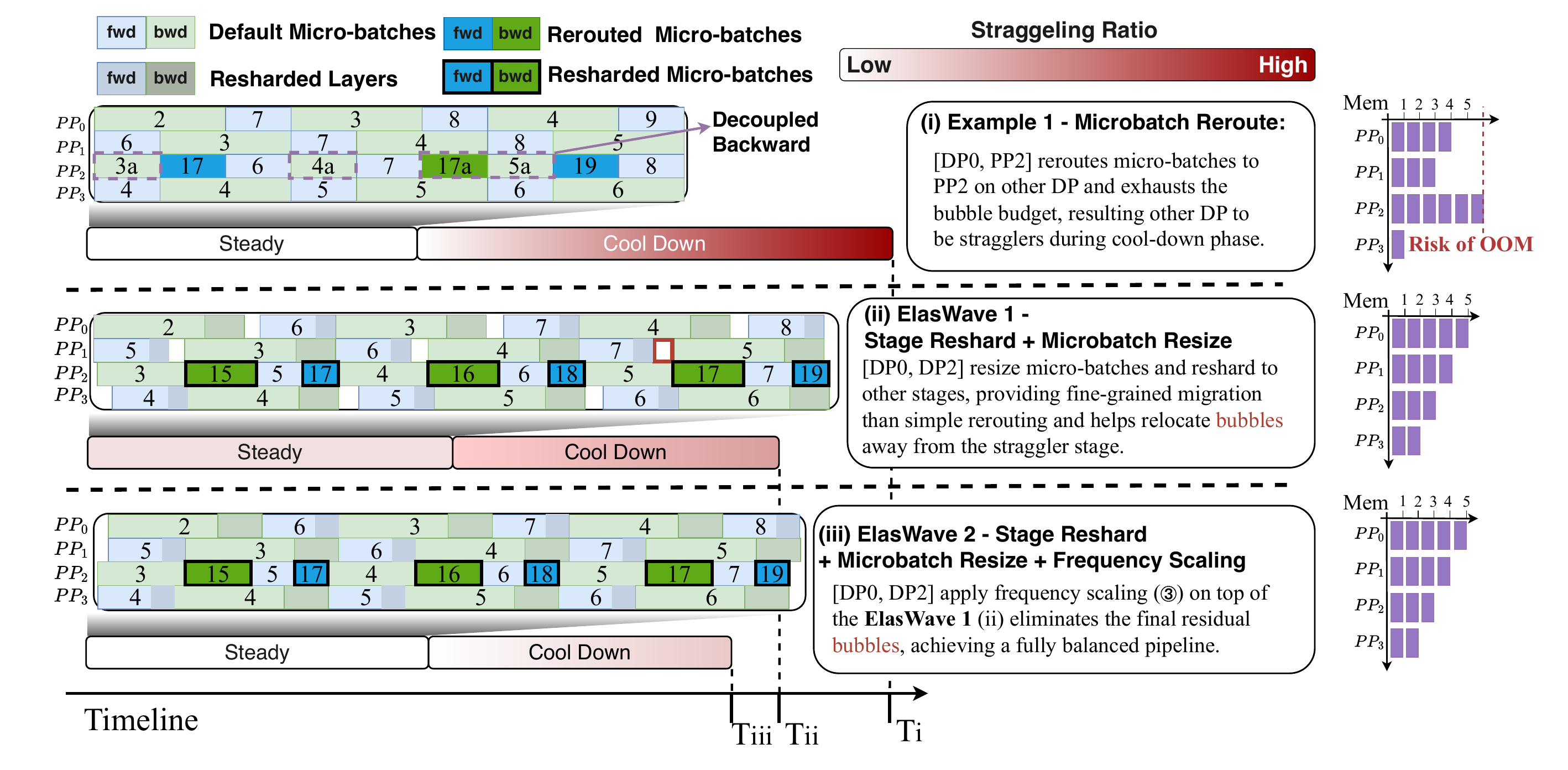}
     \caption{Comparison of pipeline schedule examples. The steady phase illustrates how multi-dim scheduling in (ii)(iii) avoids the stragglers and OOM issues in data rerouting in (i), achieving efficient and reliable training states step-by-step. The total execution times(\(T_i\), \(T_{ii}\), \(T_{iii}\)) show a progressive reduction in pipeline completion time.}
    \label{fig:multi_dimension_scheduling}
\end{figure*}

\subsection{System Context}
\name\ operates on a shared resource pool whose capacity evolves under fail-stop, fail-slow, and scheduler-driven scale changes (including preemption), which we call elastic events. Training uses hybrid parallelism with ZeRO/FSDP sharding. A Snapshot Pool maintains per-step in-memory optimizer snapshots. Elastic events can interrupt or slow down the training process. We therefore require a system that absorbs these disturbances and resumes progress while simultaneously preserving the design goals.

\subsection{System Components}
\name~is an end-to-end system with a control plane and a data plane. 
\textbf{\ding{182} Control plane} detects elastic events with \textit{{Agent}} and plans reconfiguration with \textit{Schedule engine}; 

\textit{\textbf{Agent}} is an engineering-heavy runtime co-located with each worker. It continuously monitor failures and stragglers by hooking device/host/interconnect health probes and validating liveness, while collecting memory usage, hardware utilization, and step-level performance metrics. It also listen for scheduling/preemption/resizing signals and relay them to the Core.

\textit{\textbf{Schedule engine}} governs the process control and produces a recovery plan after an elastic event. During planning, it evaluates the available device memory to ensure that the chosen actions do not exceed capacity constraints.

\textit{Dataflow planner} decides the redirection of dataflow under elasticity: it adjusts micro batch routing across pipeline stages, with the implied redistribution of activations.

\textit{Graph planner} assigns the computation graph: it repartitions the pipeline and maps partitions to hardware, and it determines the placement of ZeRO optimizer shards within each data-parallel group.

\textit{DVFS planner} selects post-event frequency settings, computing the upclock values that best remove residual bubbles while respecting power and thermal limits.

\textit{RNG planner} reshards the RNG state to remain consistent with dataflow and graph changes: it aligns per-sample/per-layer RNG usage so that randomness before and after elasticity is equivalent.

\textbf{\ding{183} Data plane}  executes recovery actions with \textit{Recovery executor} and maintains redundancy with \textit{Parameter fabric}  for rapid resumption. It executes plans with minimized MTTR and consistent parameter versions.

\textit{\textbf{Recovery executor}} execute the plan: it pauses training, sanitizes failed devices, restores communicators, reconstructs graph partitions, applies dataflow adjustments, migrates parameters, sets device frequencies, and then resumes training.

\textit{\textbf{Parameter fabric}} maintains per-step snapshots of optimizer state and redundantly backs them up across nodes; upon shrink under ZeRO, it uses these snapshots to rebuild missing shards and reestablish optimizer integrity within each data-parallel group.

\subsection{Workflow}
As illustrated in Fig. \ref{fig:System_Arch}, when the \textit{Agents} detect a failure, straggler, or scheduling signal and report a resource-pool change, the Core’s \textit{Schedule Engine} synthesizes a multi-dimensional plan—redirect dataflows, repartition the pipeline and migrate layers, set DVFS to recover high throughput, and perform RNG Resharding for computation consistency—under capacity checks. At a step boundary the \textit{Recovery Executor} briefly pauses the job, repairs connectivity by editing communicator links in place, uses the \textit{Parameter Fabric} to live-remap missing ZeRO shards from the in-memory \textit{Snapshot Pool} to ensure parameter consistency, and then applies the planned dataflow/graph/DVFS/RNG changes; communication recovery and non-blocking layer migration are optimized to minimize MTTR, after which training resumes.

%% file: Contents/5_EGS.tex
\section{Schedule Engine}\label{sec:throughput}

Elastic events change the available resource, invalidating the current training configuration. The schedule engine schedules a new training configuration based on a new resource pool along four dimensions—Dataflow, Graph, DVFS, and RNG. This coordinated plan restores a runnable setup while meeting the four targets: parameter consistency, low MTTR, high post-change throughput, and computation consistency.

\subsection{Dataflow Planner}
Upon a node failure, its micro batch dataflow (and associated activations) must be promptly rerouted to surviving nodes to sustain training progress. \name~ performs \textit{micro batch resizing} instead of micro batch number rerouting in \namerc~for multi-dimensional scheduling that avoids stragglers and OOM issues. The micro batch previously handled by the failed rank is sliced along the batch dimension evenly into DP portions and added to the surviving ranks’ micro batch sizes (Fig.~\ref{fig:rebalance}\textcircled{1}). For example, with DP $=3$ and per-rank micro batch size $=2$, a single failure yields DP $=2$ and size $=3$; the product $\text{DP}\times\text{micro batch size}$ remains constant (6), preserving the effective global batch and gradient scale.
Resizing leaves the pipeline topology and choreography unchanged and only adjusts each micro batch’s service time. Activations are produced and released by the standard forward/backward schedule, avoiding problems such as tail accumulation and extra peak memory that occur in Recycle (Figure~\ref{fig:multi_dimension_scheduling}(i)).

\subsection{Graph Planner}

The micro batch resizing policy from Dataflow Planner allows recovering at low MTTR, but applying it in isolation introduces new performance and memory hazards. When a rank fails, resizing increases the micro batch size on the affected pipeline stage; this extends that stage’s step time , turning it into the pipeline’s bottleneck. Proportionally larger micro batches also inflate activation footprints, introducing OOM risk.

\textbf{DP\&PP Domain Schedule.} The root cause is that a rank failure perturbs the global balance of work and memory across the entire pipeline. Purely local adjustments within the failed stage’s data-parallel (DP) group are therefore insufficient. Therefore, \name’s Graph Planner augments DP-domain scheduling with pipeline-parallel (PP)–domain rebalancing: in addition to resizing micro batches, it reshards PP stages to restore end-to-end balance.
As shown in Figure \ref{fig:rebalance}\textcircled{2}, \name~ allows \textit{migrating model layers} across stages to reshard stage workloads in the PP domain. By shifting a subset of layers out of the failure-impacted stage, we (i) reduce its excess compute and hence mitigate straggling, and (ii) reduce its extended activation footprint to lower the memory pressure.

\textbf{Cost model.} To determine the optimal assignment of layers to stages, we first formulate a cost model for the mini-step time on each stage. This allows us to cast the problem as a constrained minimax partition optimization. A \textit{mini-step} denotes the forward backward process of one micro batch in the pipeline. For a given stage $i$, which hosts a set of $\ell_i$ layers and processes a local micro batch of size $m_i$, the mini-step time is modeled as:

\begin{equation}
\label{eq:mini-step}
\begin{aligned}
& T_{i}^{\text{mini-step}}(\ell_i,m_i,r_{i-1},r_i,r_{i+1}) \\
&= T_{i}^{C,\mathrm f}(\ell_i,m_i) + T_{i}^{C,\mathrm b}(\ell_i,m_i) \\
&\quad + \bigl[T_{i}^{\mathrm{P2P},\mathrm f}(m_i,r_i,r_{i+1}) - \sigma_{i}^{\mathrm f} T_{i}^{C,\mathrm f}(\ell_i,m_i)\bigr] \\
&\quad + \bigl[T_{i}^{\mathrm{P2P},\mathrm b}(m_i,r_{i-1},r_i) - \sigma_{i}^{\mathrm b} T_{i}^{C,\mathrm b}(\ell_i,m_i)\bigr]
\end{aligned}
\end{equation}

Here, $T_{i}^{C,\mathrm f}$ and $T_{i}^{C,\mathrm b}$ are forward/backward compute times for stage $i$ given $\ell_i$ and $m_i$; $T_{i}^{\mathrm{P2P},\mathrm f}$ and $T_{i}^{\mathrm{P2P},\mathrm b}$ are the point-to-point activation/gradient transfer times from $i\!\to\!i{+}1$ and $i{-}1\!\to\!i$, respectively. The terms $\sigma_{i}^{\mathrm f},\sigma_{i}^{\mathrm b}\in[0,1]$ capture the overlap between compute and communication (a fraction of compute that hides communication). The quantities $r_{i-1}, r_i, r_{i+1}$ parameterize the active ranks on adjacent stages that affect achievable P2P throughput (e.g., fan-in/out or contention). In practice, $T_{i}^{C,\cdot}(\ell_i,m_i)$ are profiled offline for relevant $(\ell_i,m_i)$ pairs before training; P2P times are predicted from $m_i$ and hardware bandwidth; overlap coefficients $\sigma_{i}^{\cdot}$ are empirically profiled once and reused.

\textbf{Solver.} Given this cost model,  the planner aims to find a layer assignment that satisfies: (i) is feasible under per-stage memory capcity $\mathrm{cap}_i$ (including parameter, optimizer state, and activation memory), and (ii) minimizes the worst per-stage mini-step time, i.e.,

\begin{equation}\label{eq:minimax}
\begin{aligned}
\min_{\{\ell_i\}}\ & \max_i\ T_i^{\text{mini-step}}(\ell_i,m_i,\cdot) \\
\text{s.t.}\ & \mathrm{Mem}(\ell_i,m_i)\le \mathrm{cap}_i,\ \forall i .
\end{aligned}
\end{equation}

We solve this partitioning problem using dynamic programming over contiguous blocks of layers. Let the layers be indexed from $1$ to $L$ and stages from $1$ to $P$. For a contiguous layer block $[a..b]$ assigned to stage $p$, where $\ell_p=b-a+1$, we define its mini-step cost as $t_p([a..b]) \doteq T_p^{\text{mini-step}}(\ell_p, m_p, \cdot)$. This assignment is \textit{feasible} only if its memory footprint $\mathrm{Mem}(\ell_p, m_p)$ does not exceed the stage's capacity $\mathrm{cap}_p$.

The DP state, $f[p, \ell]$, stores the optimal minimax mini-step time for partitioning the first $\ell$ layers ($[1..\ell]$) across the first $p$ stages. The recurrence relation is formulated by choosing a split point $k \in [p-1..\ell-1]$ that partitions the layers into $[1..k]$ and $[k+1..\ell]$, minimizing the maximum cost for the two subproblems:
\begin{equation}
    f[p, \ell] = \min_{k \in [p-1..\ell-1]} \max\{ f[p-1, k], t_p([k+1..\ell]) \}
    \label{eq:dp_recurrence}
\end{equation}
After computing the DP table, we backtrack to find the optimal stage boundaries $\{b_1, \dots, b_{P-1}\}$. The complete algorithm is detailed in Alg.~\ref{alg:minimax-dp}.

The DP approach first guarantees memory feasibility (no OOM errors at any stage). and then optimizes the minimax objective, which directly targets pipeline throughput. To enable rapid decision-making at failure time, all required segment costs ($\mathrm{Mem}[u..v]$ and $t_p([u..v])$) are precomputed from offline profiles and bandwidth models. While the theoretical complexity is $O(P\, L^2)$, the solver is efficient in practice due to aggressive pruning of infeasible partitions and the small number of stages ($P$) typical in pipeline parallelism.

By coupling micro batch resizing with layer re-partitioning under a principled minimax objective, our planner restores balance across both computation and memory and thereby recovers pipeline throughput while maintaining low MTTR.

\algnewcommand\algorithmicnotation{\textbf{Notation:}}
\algnewcommand\Notation{\item[\algorithmicnotation]}
\begin{algorithm}[t]
\caption{Minimax Layer Partition (DP)}
\label{alg:minimax-dp}
\begin{algorithmic}[1]\small
\Require $L$ = \#layers;\ $P$ = \#stages;\ per-stage memory caps $\mathrm{cap}[1..P]$;\
memory segment cost $\mathrm{Mem}[u..v]$;\ mini-step cost $t_p([a..b])$ = $T_p^{mini-step}(b-a)$
\Ensure $f[P,L]$ = optimal worst-stage mini-step time over all feasible $P$-way partitions;\
$\{b_1,\dots,b_{P-1}\}$ = right boundaries (stage $j$ runs $(b_{j-1}{+}1)..b_j$, with $b_0{=}0,b_P{=}L$)
\Notation $k$ = split index;\ $k^{\ast}(p,\ell)$ = optimal split for state $(p,\ell)$

\For{$\ell=1..L$}
     \State $f[1,\ell]\leftarrow t_1([1..\ell])$
\EndFor

\Statex \Comment{Transition}
\For{$p=2..P,\ \ell=p..L$}  
  \State $\begin{aligned}[t]
    k^{\ast}(p,\ell) &\gets \operatorname*{arg\,min}_{k\in[p-1,\ell-1]}
        \max\{\,f[p-1,k],\ t_p([k{+}1..\ell])\,\}\\
    \text{s.t.}\quad & \mathrm{Mem}[k{+}1..\ell]\le \mathrm{cap}[p]
  \end{aligned}$
  \State $f[p,\ell]\gets \max\{\,f[p-1,k^{\ast}(p,\ell)],\ t_p([k^{\ast}(p,\ell){+}1..\ell])\,\}$
\EndFor

\State $b_{P-1}\leftarrow k^{\ast}(P,L)$;\quad
for $p=P{-}1$ down to $2$:\ $b_{p-1}\leftarrow k^{\ast}(p,b_p)$

\State \Return $f[P,L]$ and $\{b_1,\dots,b_{P-1}\}$
\end{algorithmic}
\end{algorithm}

\subsection{DVFS Planner}

Layer migration in the PP domain (Graph Planning) restores balance at the granularity of layers. However, the granularity of a layer can be coarser than the residual performance gap among stages after rebalancing. In such cases, any further layer migration would overshoot: the recipient stage would become the new straggler because a whole layer’s worth of work exceeds the remaining bubble. This situation is illustrated in Fig. \ref{fig:multi_dimension_scheduling}(b): the failure-impacted stage still exhibits mild straggling, yet moving another layer would invert the bottleneck. To eliminate these sub-layer-scale bubbles, we complement DP- and PP-domain scheduling with compute-unit scheduling via dynamic voltage and frequency scaling (DVFS).

Our policy is to up-clock only the residual straggler stage to shorten its mini-step time until it aligns with its peers, thereby removing the remaining bubble without perturbing the pipeline assignment. Because sustained high frequencies may accelerate hardware aging, we aim for the minimum necessary frequency increase. 

\textbf{Solver.} Because higher frequency can stress hardware, we aim for the minimum necessary uplift. The controller proceeds in two steps (Alg. \ref{alg:dvfs-alg}). First, it tests feasibility by setting the straggler to $f_{\max}$ and measuring its mini-step over a short observation window $W$. If even at $f_{\max}$ the stage still lags the target $T^\star$ (within tolerance $\varepsilon$), the gap is not compute-bound and is marked UNACHIEVABLE. Otherwise, alignment is possible; the controller runs a simple bisection between the current frequency and $f_{\max}$ to find the lowest frequency that meets $T^\star$ (respecting a minimum step $\Delta f_{\min}$).

\textbf{Bubble-free restoration.} By combining DP-domain resizing, PP-domain layer migration, and DVFS up-clocking, \name~removes the bubbles introduced by node failures across the pipeline. As shown in Fig. \ref{fig:multi_dimension_scheduling}(c), the pipeline returns to a bubble-free status to maximize the throughput.

\begin{algorithm}[b]
\caption{Minimum Bisection Frequency Scaling}
\label{alg:dvfs-alg}
\small
\begin{algorithmic}[1]
\Require current frequency \(f_{\mathrm{cur}}\), maximum \(f_{\max}\), target \(T^\star\), tolerance \(\varepsilon\), minimum step \(\Delta f_{\min}\), Observation window \(W\)
\Ensure \((f^\star, \textsc{status})\) with \(\textsc{status}\in\{\textsc{ACHIEVABLE}, \textsc{UNACHIEVABLE}\}\)
\State \(t_{\mathrm{cur}} \gets \textsc{obs\_time}(W)\)
\If{\(|t_{\mathrm{cur}}-T^\star|\le \varepsilon\)} \Return \((f_{\mathrm{cur}}, \textsc{ACHIEVABLE})\) \EndIf
\State \textsc{apply\_freq}\((f_{\max})\);\quad \(t_{\max}\gets \textsc{obs\_time}(W)\) 
\If{\(t_{\max} > T^\star+\varepsilon\)} \Return \((f_{\max}, \textsc{UNACHIEVABLE})\) \EndIf
\State Define evaluator \(\mathcal{E}(f):\ \textsc{apply\_freq}(f);\ \textsc{obs\_time}(W)\le T^\star+\varepsilon\)
\State \(f^\star \gets \textsc{bisect\_min\_feasible}\big(f_{\mathrm{lo}}{=}f_{\mathrm{cur}},\, f_{\mathrm{hi}}{=}f_{\max},\, \mathcal{E},\, \Delta f_{\min}\big)\)
\State \Return \((f^\star, \textsc{ACHIEVABLE})\)
\end{algorithmic}
\end{algorithm}

\subsection{RNG Resharding}
\begin{figure}[t]
  \centering
  \begin{subfigure}{\linewidth}
      \centering
      \includegraphics[width=0.9\linewidth]{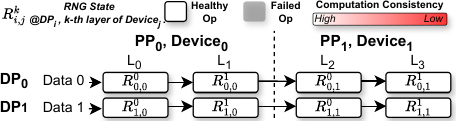}
      \vspace{-5pt}
      \caption{Original training with even layer distribution}
  \end{subfigure}
  \begin{subfigure}{\linewidth}
      \centering
      \includegraphics[width=0.9\linewidth]{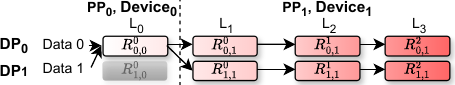}
      \vspace{-5pt}
      \caption{Elastic training w/ ElasWave, w/o RNG Resharding}
  \end{subfigure}
  \begin{subfigure}{\linewidth}
      \centering
      \includegraphics[width=0.9\linewidth]{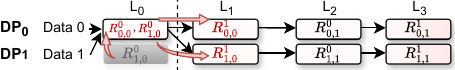}
      \vspace{-5pt}
      \caption{Elastic training w/ ElasWave, w/ RNG Resharding}
  \end{subfigure}
  \vspace{-10pt}
  \caption{Elastic training without (b) or with (c) RNG Resharding. In (b), $L_1$ in $PP_0$ will be transferred to $PP_1$, the RNG state $R_{0,1}^0$ and $R_{1,1}^0$ in $PP_1$ will be directly applied to $L_1$, which introduces inconsistency. Besides, Data 1 is allocated to $DP_0$, but will be processed by RNG state $R_{0,0}^0$, which is also inconsistent with (a). In (c), RNG state $R_{1,0}^0$ will also be saved in $DP_0$, and will be applied to Data 1. After processing $L_0$, $R_{0,0}^1$ and $R_{0,1}^1$ will be sent to $PP_1$ of $DP_0$ and $DP_1$, respectively, and be used to process the trasferred $L_1$. Therefore, (c) achieves a similar convergent behavior.}
  \label{fig:rng_resharding}
\end{figure}

To improve numerical consistency of elastic training, we propose the RNG (Random Number Generator) Resharding for random operations, such as dropout. 
In the distributed training, each node has an RNG to generate random number for random operations. As Figure~\ref{fig:rng_resharding} (b) shows, both layer rebalance and micro batch rebalance can change the RNG state for a specific data, which will further change the consistent behavior. Therefore, we propose the RNG Resharding in Figure~\ref{fig:rng_resharding} (c), which contains two steps for the RNG consistency.

In the layer rebalance, several layers in the failed stage will be transferred to other stages to avoid the affect of stragglers. Correspondingly, we need to transfer the RNG state from the failed stage in every forward propagation, and only apply the transferred RNG state to the transferred layers.

In the micro batch rebalance, data in the failed node will be dispatched to other nodes in the same stage. The dispatched data are supposed to be processed with their original RNG state. Therefore, every node needs to backup all RNG state of other nodes in the same stage, and uses corresponding RNG state to process the dispatched data. 

In this way, we can ensure all data are processed with the same RNG state as that in the normal training, and achieve a same convergent behavior. Besides, we adjust the computation of average gradient in the global batch, so that the unevenly divided micro batch will not affect the final gradient results. We also notice that the change of the float-point addition order may introduce a small difference in the elastic schedule. However, in our evaluation in Section \ref{subsec:exp_conv} , we find that it will not affect convergence consistency or cause loss spikes.

%% file: Contents/4_ZCE.tex
\section{Parameter Fabric: Snapshot \& Live Remap}

\subsection{Per-step Snapshot}
While ZeRO's memory efficiency makes it a default for large model training, its partitioned-state design fundamentally conflicts with elastic scaling by removing data redundancy. Our per-step snapshot mechanism provides fault tolerance for ZeRO-based training with minimal overhead. 

\begin{figure}[t]
  \centering
  \begin{subfigure}{\linewidth}
    \centering
    \includegraphics[width=.95\linewidth]{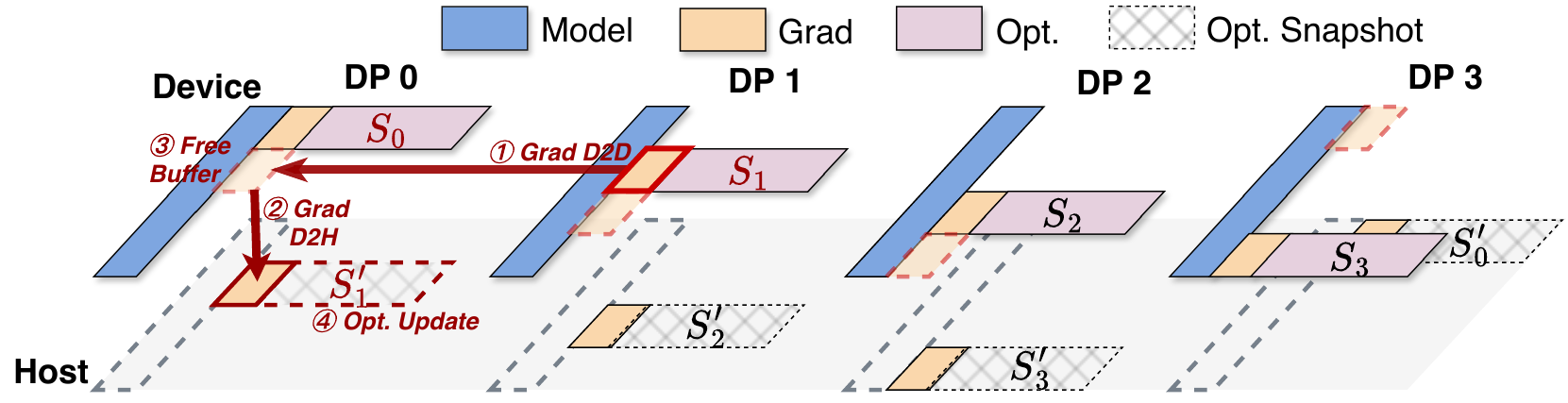} %
    \caption{Data Flow: Each worker (e.g., DP 1) sends the gradient shard for its local partition ($S_1$) to its peer (e.g., DP 0) via D2D (\ding{172}). DP 0 then offloads this gradient to host memory (\ding{173} D2H), frees the device buffer (\ding{174}), and its host CPU updates the corresponding snapshot optimizer state on the host ($S'_1$) (\ding{175}).}  
    \label{fig:ckpt_a}
  \end{subfigure}

  \par\medskip 

  \begin{subfigure}{\linewidth}
    \centering
    \includegraphics[width=.95\linewidth]{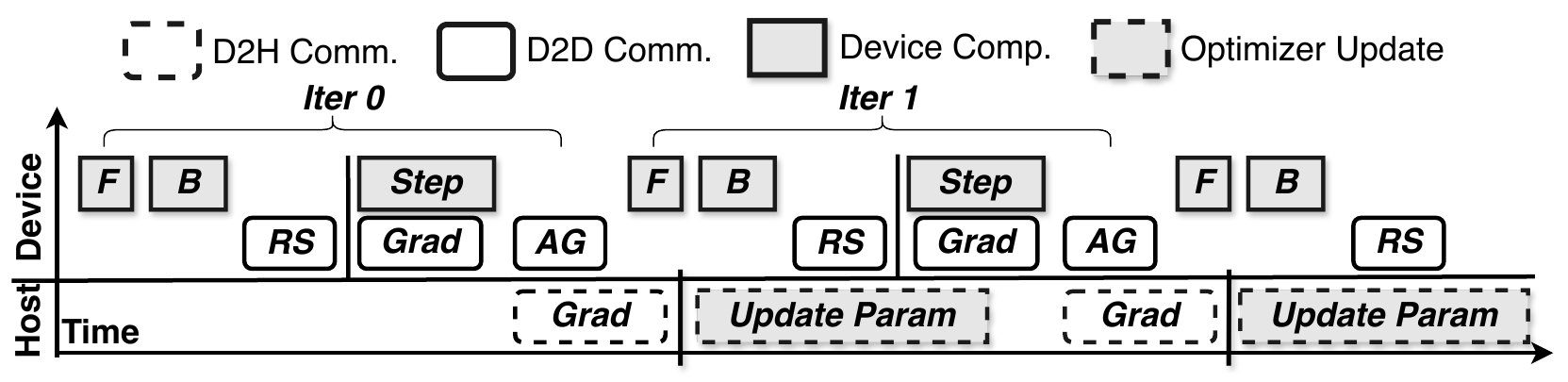} %
    \caption{Timeline: After Forward (F), Backward (B), and Reduce-Scatter (RS) passes, the D2D gradient transfer (Grad) for snapshotting occurs, running parallel to the optimizer update (Step). The gradient is then offloaded to the host (D2H Grad), overlapping with All-Gather (AG). The host's parameter update (Update Param) is hidden by the next iteration's computation, keeping the critical path clear.}
    \label{fig:ckpt_b}
  \end{subfigure}

  \caption{The asynchronous per-step snapshot mechanism.}
  \label{fig:ckpt}
\end{figure}

In a standard ZeRO setup, each worker's GPU holds a unique partition of the optimizer states, which we denote as $O_i^{\text{device}}$. To introduce redundancy, we implement a ring-based snapshot scheme, as shown in Figures~\ref{fig:ckpt_a}. Each worker $i$ becomes responsible for backing up the optimizer state partition from its neighbor, worker $(i+1) \bmod n$. This snapshot, denoted as $O_i^{\text{host}}$, is stored in worker $i$'s host memory.

The design for achieving minimal overhead is illustrated in Figure~\ref{fig:ckpt_b}. The key principle is to make the snapshot process asynchronous and communication-efficient. Instead of transferring bulky optimizer states, we only transmit compact gradient shards to a peer worker, reducing snapshot communication by at least 4x for a mixed-precision optimizer like Adam. The actual parameter update for the snapshot is offloaded to the host CPU. By decoupling the resources (device vs. host) and execution timelines, the entire snapshot operation is performed in the background and overlaps with the next training iteration's computation. This ensures fault tolerance with negligible performance impact, as the critical path of training is not stalled.

\begin{figure}[htbp]
  \centering

  \begin{subfigure}{\linewidth}
    \centering
    \includegraphics[width=.95\linewidth]{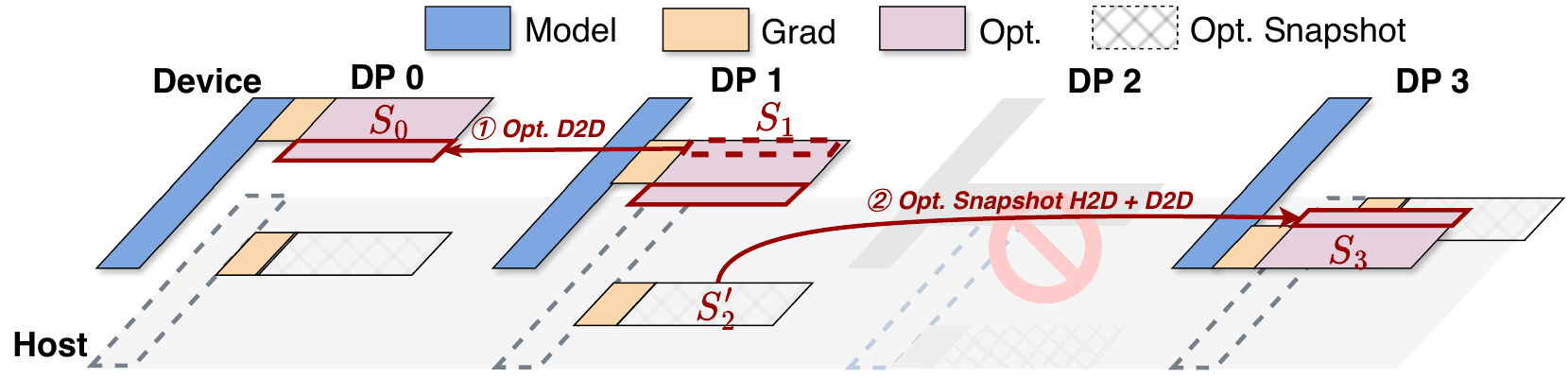}
    \caption{Data Flow: When worker DP 2 fails, its state is recovered from snapshots on other workers (e.g., DP 1). Guided by the overlap matrix ($M_{\text{overlap}}$), worker DP 0 pulls necessary shards from DP 1's device via D2D (\ding{172}), while DP 3 retrieves its required shards from DP 1's host snapshot ($S'_1$) through H2D+D2D (\ding{173}).}
    \label{fig:scale_down_a}
  \end{subfigure}

  \par\medskip

  \begin{subfigure}{\linewidth}
    \centering
    \includegraphics[width=\linewidth]{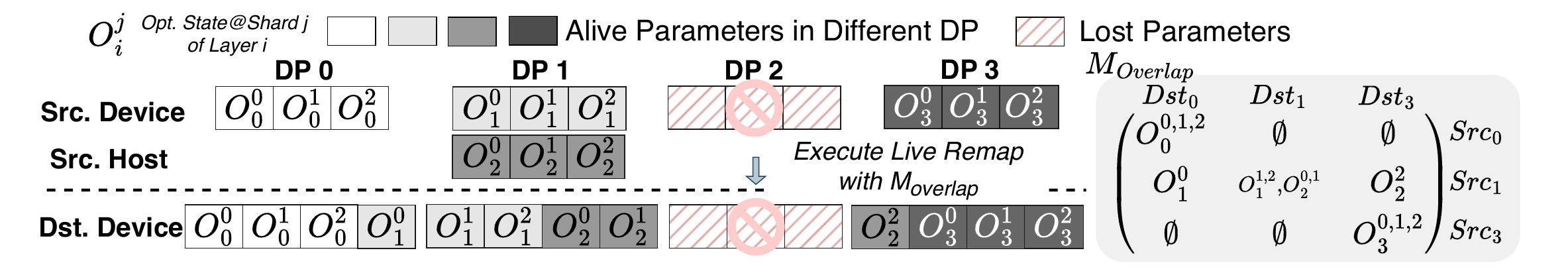} 
    \caption{Remap Plan ($M_{\text{overlap}}$): The matrix defines the data transfer plan. For instance, the entry at (Src$_1$, Dst$_0$) indicates that partition $O^0_1$ must be transferred from worker 1 to worker 0, while diagonal entries like (Src$_0$, Dst$_0$) represent data that remains local.}
    \label{fig:scale_down_b}
  \end{subfigure}

  \caption{Resharding process for a scale-down event.}
  \label{fig:scale_down}
\end{figure}

\subsection{Live Remap}
Live Remap orchestrates the redistribution of optimizer states in response to any scaling event.
Upon any scaling event, Live Remap initiates a four-step process. First, for scale-downs, an \textit{\ding{172}Integrity Check} (Figure~\ref{fig:scale_down_a}) identifies failed workers (e.g., $F = \{2\}$) and confirms their state is recoverable from remaining on-device ($O_i^{\text{device}}$) and snapshot ($O_i^{\text{host}}$) partitions. Next, the system \textit{\ding{173} Computes a Transfer Plan}. It creates the consolidated partitions ($O_{i, \text{consolidated}}$) as the logical union of all available on-device and host-snapshot data. It then computes an overlap matrix ($M_{\text{overlap}}$) by intersecting these source partitions with the final target partitions ($O_{j, \text{target}}$), defining the precise data flow (Figure~\ref{fig:scale_down_b}). The plan is executed in \textit{\ding{174} Opt. Redistribution} via D2D and H2D communication. Finally, in \textit{\ding{175} Finalization}, workers reconstruct new states and free unused memory.

%% file: Contents/6_CRR.tex
\section{MTTR Minimization} 

\subsection{Dynamic Communicator}
Modern training jobs face frequent hardware failure, but traditional distributed frameworks assume fixed resources and static communication domains, thus often force costly communicator reinitialization, incurring high overhead and degraded performance under resource change. NCCL exposes a communicator shrink API, but it essentially rebuilds a new communicator from the surviving ranks, still paying heavy reinitialization cost \cite{bachan_nccl227_2025}.

\begin{figure}[t]
  \centering
  \includegraphics[width=0.4\textwidth]{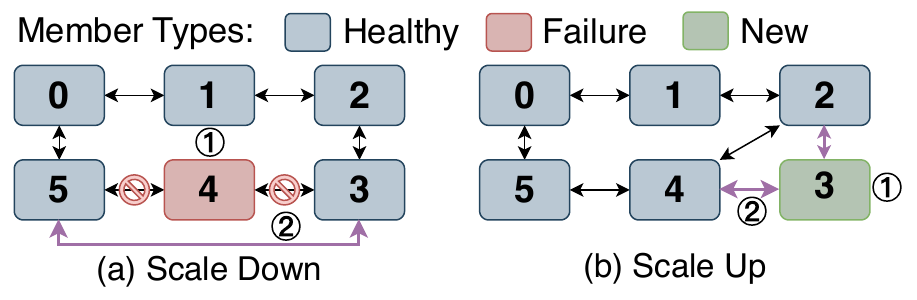}
  \caption{Dynamic Communicator Operations: 
  \normalfont{(1) Scale-down: When an error is detected on a rank/node, it is  removed from the worker pool to allow for training to continue, and only affected communicators are adjusted, while reusing existing links. 
  (2) Scale-up: When a new worker joins the pool, only specific communicators are adjusted, and existing links are reused.
  }}
  \label{fig:DynamicCommunicatorOperations}
\end{figure}

To address this limitation, we introduce the \textit{Dynamic Communicator}, an adaptive framework that \textit{scales up or scales down} (Figure~\ref{fig:DynamicCommunicatorOperations}) in real time to respond to changes in resource allocation by smoothly adapting communication interfaces. We adapt communicators online by \emph{reusing existing links} and modifying only the affected groups. When resources change, the system creates only missing connections and preserves intact ones, avoiding global rebuilds and enabling recovery without a full restart.

\textbf{Scalability.}
This approach provides (i) adaptive communication management for scale-up/down, (ii) efficient failure handling w/o full restarts, and (iii) efficient link management by creating only missing connections during dynamic operations.
\ding{182} \emph{Scale-down:} upon detecting a failed rank, the worker is removed and neighbors are reconnected to survival links (\textcircled{1}), while only the necessary local communicators are updated (\textcircled{2}).
\ding{183} \emph{Scale-up:} when a new worker joins, it establishes just the additional links (\textcircled{1}–\textcircled{2}); existing links and unrelated communicators are reused.

This \emph{in-place, incremental} optimization eliminates cluster-wide rebuilds, \emph{flattening the recovery cost with respect to communication scale} and rendering \emph{MTTR a constant, sub-second bound}.

\subsection{Model Recovery Acceleration}\label{sec:async_layer}
\begin{figure}[b]
  \centering
  \includegraphics[width=\linewidth]{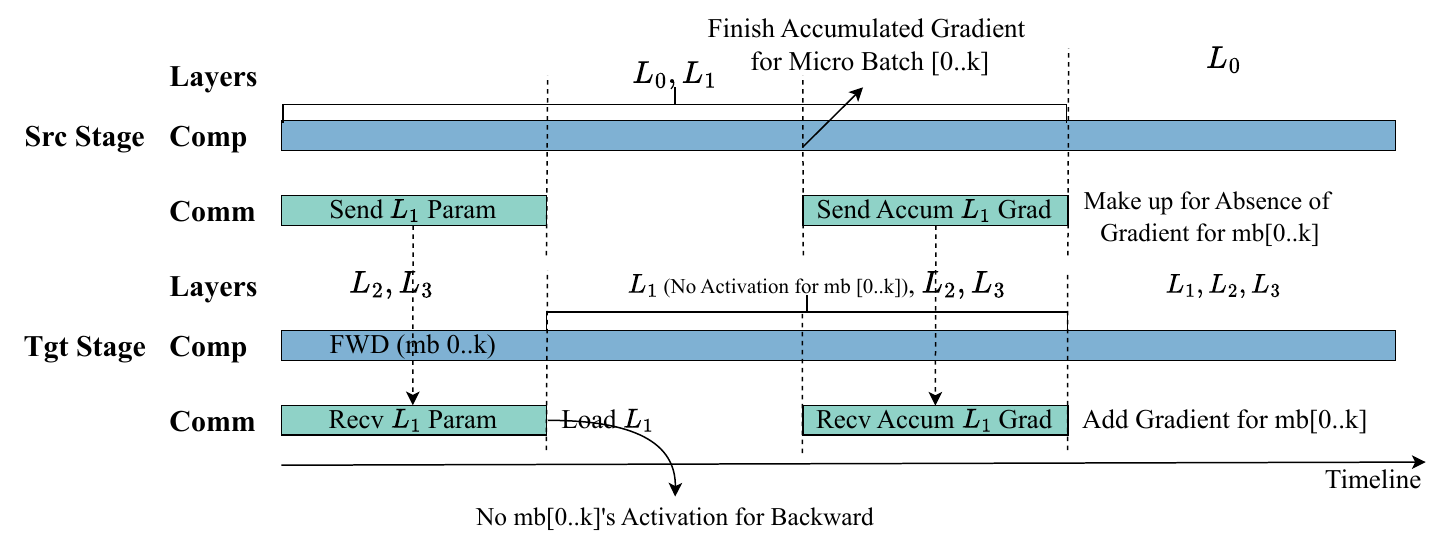}
  \caption{ Asynchronous layer migration with gradient precomputation. While $L_1$ parameters stream to the target, the target proceeds with forward for micro-batches $0..k$ through its other layers, so no $L_1$ activations exist for their backward. The source keeps a shadow instance of $L_1$, completes backward for $0..k$, accumulates the missing $L_1$ gradients, and asynchronously sends them to the target, which merges them to obtain a complete gradient without blocking.
}
  \label{fig:layer-transfer}
\end{figure}

A straightforward method is \textbf{Blocked Layer Migration}, which copies the migrating layer to its new stage, after which resumes training. The stall scales with payload and bandwidth; when moves are frequent or span multiple layers, these stalls accumulate directly into MTTR.

\textbf{Async. resharding with gradient precomputation.}
Overlapping the copy with training avoids the stall, but if the target processes micro-batches \textit{before} the layer arrives, it cannot contribute that layer’s gradients for those micro-batches, breaking gradient accumulation and forcing a later pause.
Our method keeps the overlap and preserves accumulation (Fig. \ref{fig:layer-transfer}). While the target proceeds, the source runs a shadow instance of the migrating layer for the early micro-batches (mb[0..k]), accumulates their missing gradients, and asynchronously ships this “payback” gradient to the target. The target merges it with local contributions once parameters are loaded. The only added cost is one gradient transmission per move, scheduled at lower priority and overlapped with ongoing compute, yielding non-blocking migration with complete gradient accumulation.

\subsection{ZeRO Optimizer Recovery Acceleration}

\begin{figure}[t]
  \centering

  \begin{subfigure}{\linewidth}
    \centering
    \includegraphics[width=0.9\linewidth]{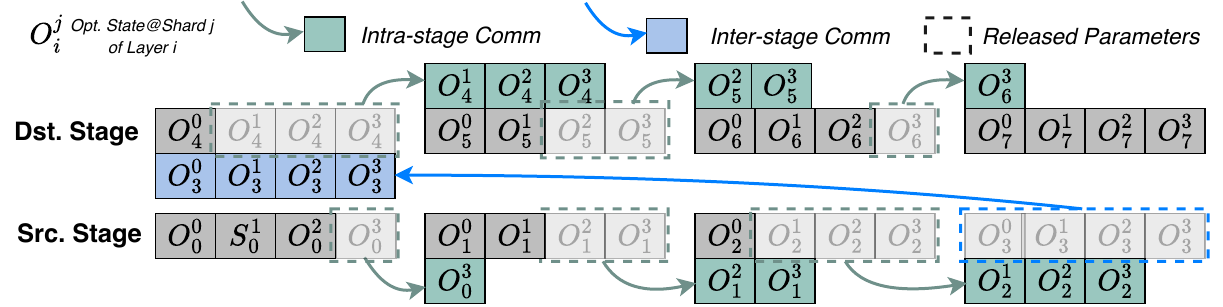}
    \caption{Contiguous: When a layer's optimizer state moves, its constituent shards ($O_i^j$, the $j$-th shard of layer $i$'s state) are transferred from the source (down) to the destination (up). This triggers an all-to-all re-sharding within both stages, shown by green arrows, to restore a contiguous layout for the remaining shards.}
    \label{fig:b}
  \end{subfigure}\par\vspace{6pt}
  \begin{subfigure}{\linewidth}
    \centering
    \includegraphics[width=0.9\linewidth]{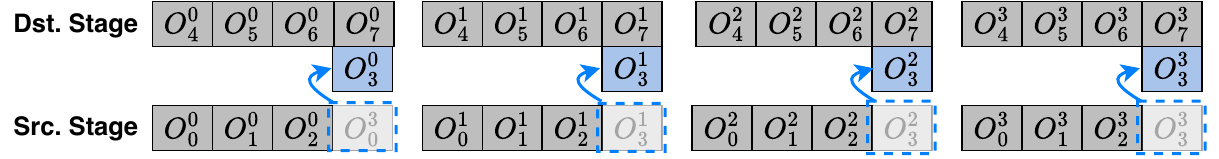}
      \caption{Interleaved: The optimizer state is partitioned such that rank $j$ owns the $j$-th shard ($O_i^j$) for every layer. Migration thus reduces to direct point-to-point transfers, where each rank $j$ sends its shard to the corresponding rank $j$ in the destination stage, eliminating any intra-stage re-sharding.}
    \label{fig:c}
  \end{subfigure}

  \caption{Comparison of optimizer parameter migration under Contiguous vs. Interleaved ZeRO assignments.}
  \label{fig:optimizer-transfer}
\end{figure}

\textbf{Contiguous Assignment and Intra-Stage Resharding.}
Migrating the optimizer state $O_i$ of layer $i$ from pipeline stage $S$ to $S{+}1$ under the Contiguous assignment triggers all-to-all(v) resharding within both the source and destination DP groups, which dominates migration time. In this layout, each DP group maintains a single global byte array, and the ownership invariant requires each rank to hold one contiguous block of approximately equal size. After exporting $O_i$ (Fig. \ref{fig:optimizer-transfer}a), the new cut points shift by $\approx |O_i|/D$ across the group, so multiple original intervals overlap each target interval; restoring contiguity therefore requires many-to-many personalized exchanges across ranks. In the figure, green arrows denote intra-stage exchanges and the dashed regions indicate released bytes once the layer has moved.

\textbf{Interleaved Assignment and Point-to-Point Migration.}
Under the Interleaved assignment, migration reduces to $D$ disjoint rank-to-rank sends of the layer’s shards and eliminates any intra-stage resharding. Each layer is uniformly partitioned so that DP rank $j$ always owns $O_i^j$ for every layer. Consequently, moving $O_i$ from stage $S$ to $S{+}1$ consists only of sending $O_i^j$ from rank $j$ at stage $S$ to rank $j$ at stage $S{+}1$ (Fig. \ref{fig:optimizer-transfer}b); no stage-internal reshaping is needed.

\textbf{Communication Cost.}
The Contiguous migration comprises a cross-stage transfer of $|O_i|$ plus intra-stage resharding that can be executed in $D{-}1$ neighbor rounds with cost $\tfrac{D-1}{2}\,|O_i|$. The total is therefore
$
\ \tfrac{D+1}{2}\,|O_i| \ \text{bytes}.
$ In contrast, Interleaved performs exactly $D$ 1:1 sends summing to
$
\ |O_i| \ \text{bytes}.
$

The change of ZeRO is purely an ownership-layout transformation: optimizer semantics and updates are unchanged, and each $O_i$ is reconstructed from its shards exactly as in standard ZeRO.

%% file: Contents/8_Evaluation.tex
\section{Evaluation}
\subsection{Experimental Setup}
We conduct experiments on a 12-node Ascend cluster, where each node has 8× Ascend 910B NPUs (32 GB memory) connected via 200 Gbps RoCE links; the software stack is CANN 8.0RC.3; the NPUs' initial frequency is 1,400 MHZ, with a maximum frequency of 1,650 MHZ. We compare \name~w\allowbreak ith two state-of-the-art systems: \nametft, a widely-used industrial solution with DP-replica granularity elasticity, where an entire replica is removed on failure; and \namerc, the academic state-of-the-art using data rerouting, which re-routes micro-batches within a DP domain to fit into pipeline bubbles and avoid straggling. However, vanilla 1F1B has become a weak baseline due to its low MFU and large bubble size. For fair comparison, we choose an SOTA pipeline on 1F1B: AdaPipe~\cite{10.1145/3620666.3651359}, which finds an optimal initial layer distribution to create a bubble-less schedule with maximized MFU. Our workloads consist of three Llama 2 models, with detailed configurations in Table~\ref{tab:workload-configs}. 

\begin{table}[htbp]
\centering
\caption{Workload Configurations for Llama 2 Models.}
\label{tab:workload-configs}
\setlength{\tabcolsep}{4pt} 
\resizebox{1\columnwidth}{!}{
\begin{tabular}{l c c c}
\toprule
Model & \begin{tabular}[c]{@{}c@{}}Parallelism\\ (TP,PP,DP)\end{tabular} & \begin{tabular}[c]{@{}c@{}}Micro-batch\\ Size\end{tabular} & \begin{tabular}[c]{@{}c@{}}Global Batch\\ Size\end{tabular} \\
\midrule
Llama2-7B  & (4, 3, 8) & 4 & 8192 \\
Llama2-13B & (4, 6, 4) & 2 & 2048 \\
Llama2-34B & (4, 8, 3) & 1 & 768  \\
\bottomrule
\end{tabular}
}
\end{table}

\subsection{Throughput Under Fail-stop Failures}
\begin{figure}[!htbp]
  \centering
  \includegraphics[width=\columnwidth]{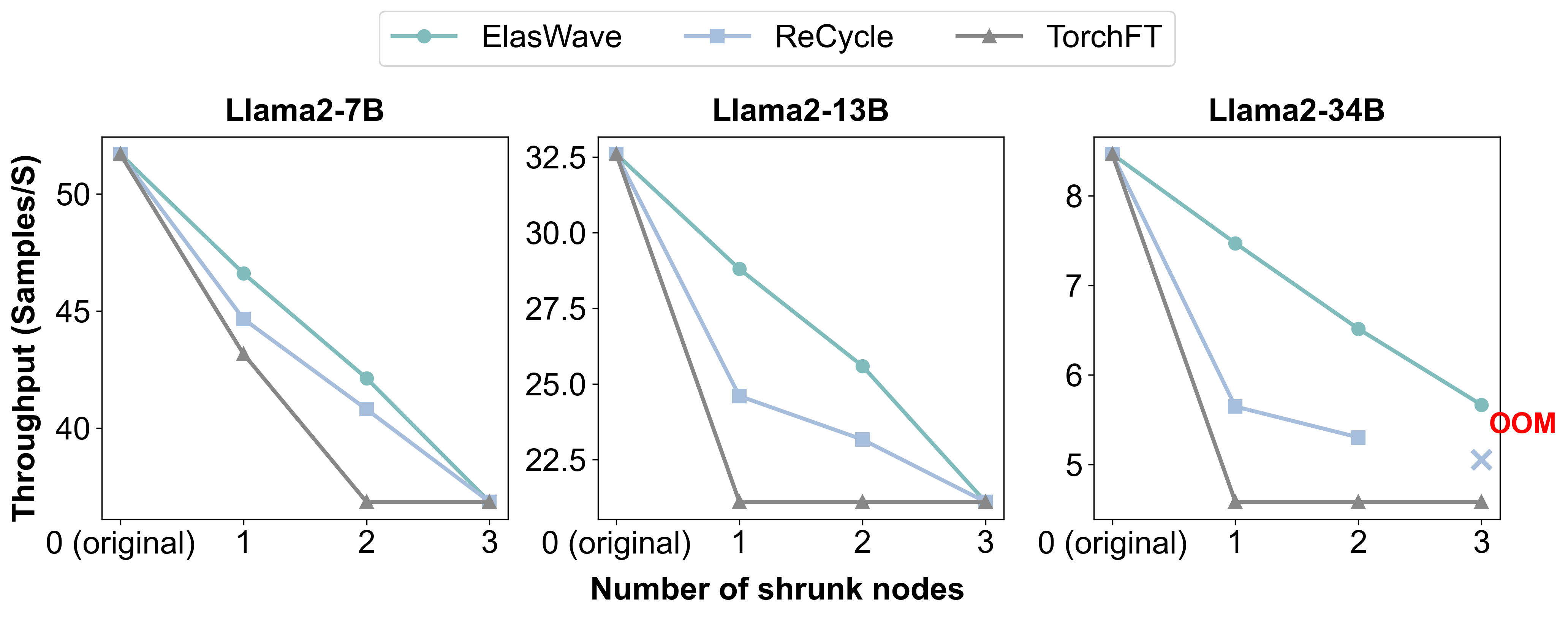}
  \caption{Throughput comparison under fail-stop failures.}
  \label{fig:failure-throughput}
\end{figure}
Across models and shrink magnitudes, throughput orders \name~> \namerc~> \nametft~(Fig. 11). \nametft~is worst because each shrink drops whole DP replicas, yielding idle capacity and cliff-like losses. \namerc~reroutes all failed work within a single PP stage; with many micro-batches the bubble budget is insufficient, creating stage stragglers—hence a sharp drop at the first shrink, smaller additional losses thereafter, and an OOM at Llama2-34B with three-node loss due to deferred weight-gradient memory. \name~spreads the failed load globally via graph migration, so throughput degrades nearly linearly. For example, on Llama2-34B with one node shrink it shows ~60\% higher throughput than \nametft~and ~35\% higher than \namerc. When the lost NPUs equal an integer multiple of the DP-replica size, \namerc~and \name~degenerate to \nametft~(e.g., Llama2-7B at 3 nodes shrink, Llama2-13B at at 3 nodes shrink).

\begin{figure}[!htbp]
    \centering
    \begin{subfigure}{0.48\linewidth}
        \centering
        \includegraphics[width=\linewidth]{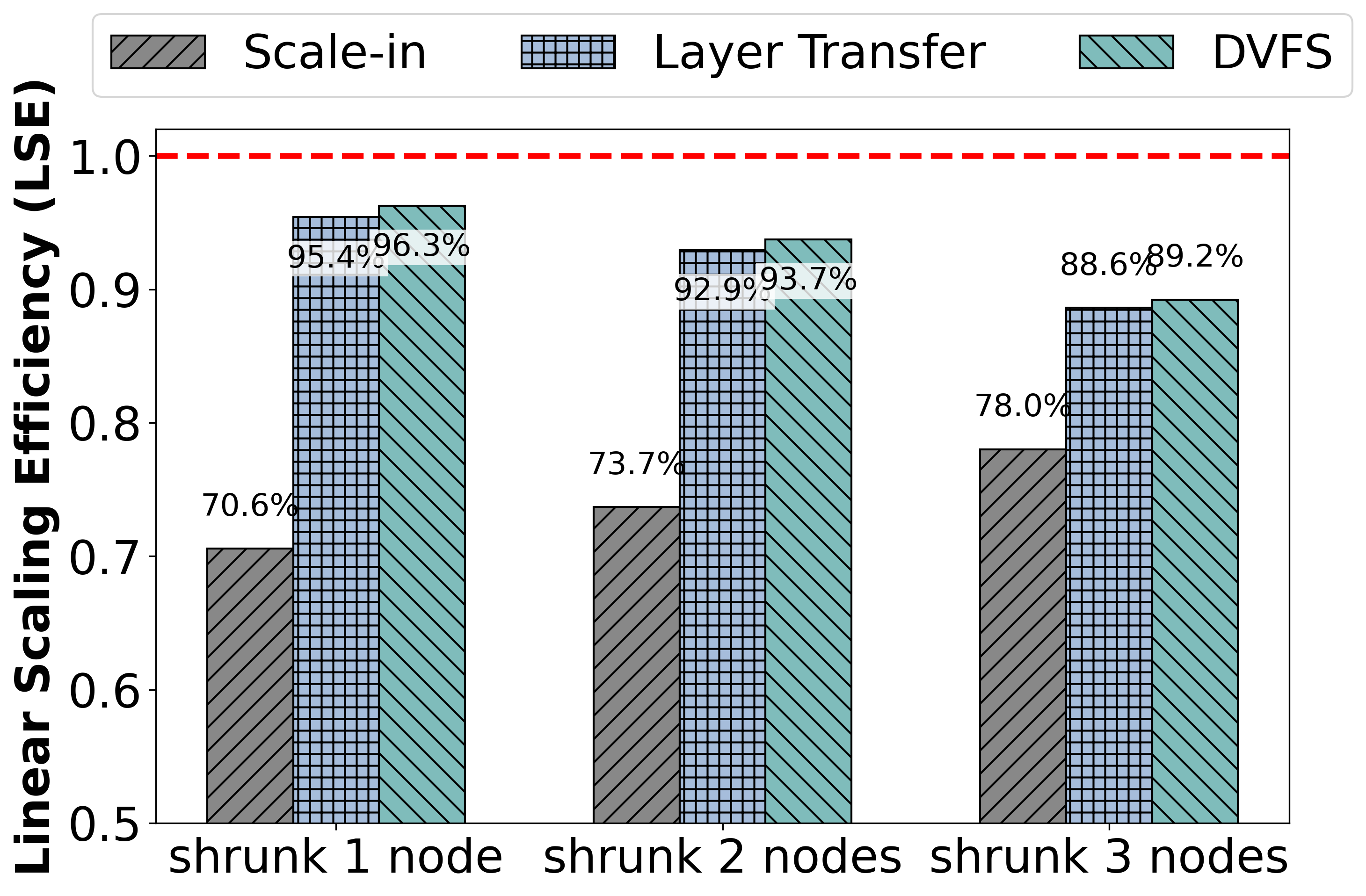}
        \caption{Throughput efficiency (LSE) gains from each optimization under 1, 2, and 3 node failures.}
        \label{fig:throughput-breakdown}
    \end{subfigure}
    \hfill
    \begin{subfigure}{0.48\linewidth}
        \centering
        \includegraphics[width=\linewidth]{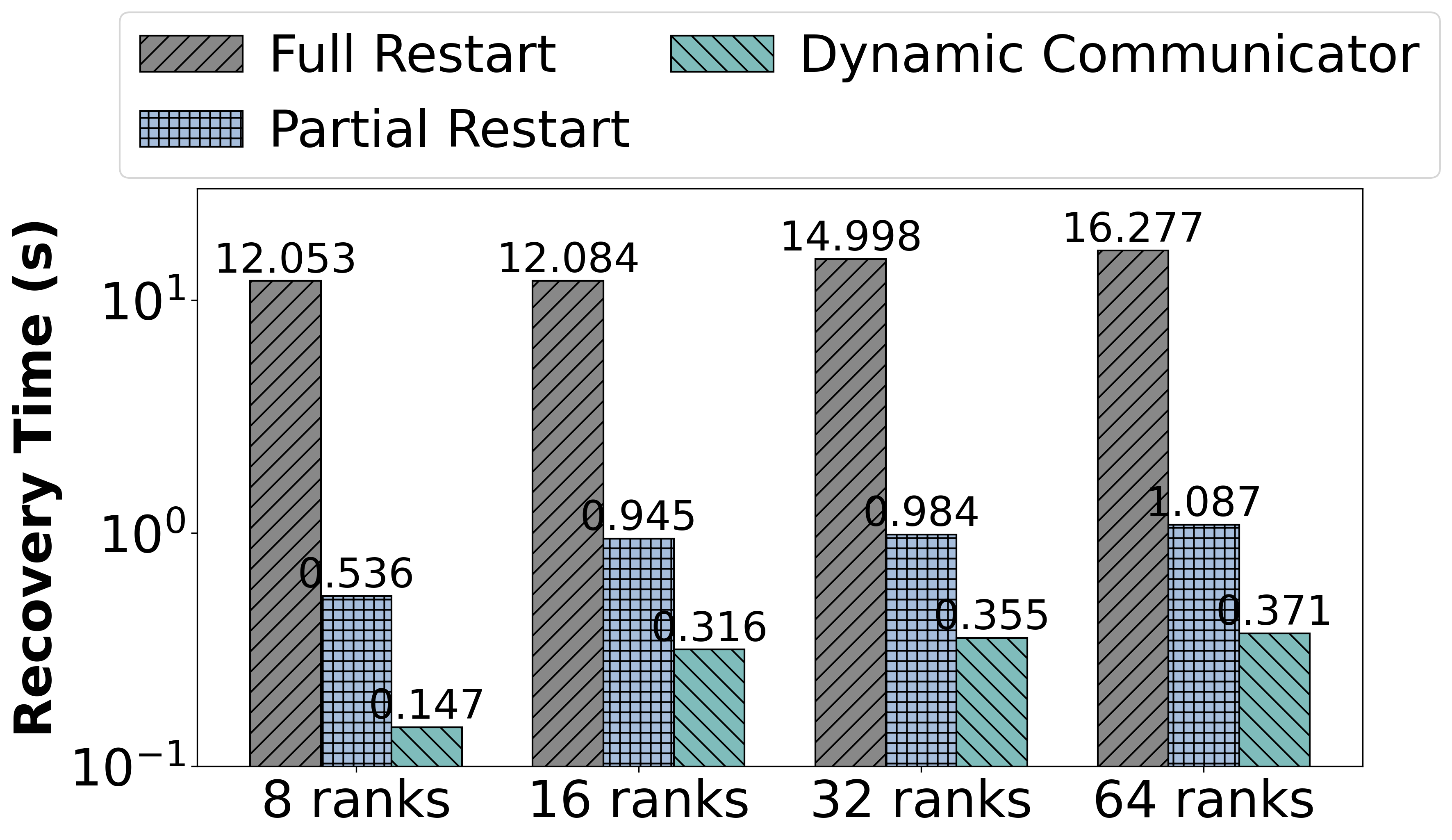}
        \caption{MTTR for three recovery scenarios: Full Restart (Grey), Partial Restart (Blue), and our Dynamic Communicator (Green).}
        \label{fig:comm-group-update}
    \end{subfigure}
    \caption{\name's performance under failures, showing (a) the breakdown of throughput efficiency gains and (b) the communication group recovery time (MTTR).}
    \label{fig:performance-summary}
\end{figure}


\textbf{Throughput Breakdown.} We use Linear Scaling Efficiency (LSE) to attribute gains across \name’s optimization steps. LSE serves as a “performance score,” where a higher value indicates that throughput degradation upon node loss is closer to the ideal linear scaling, signifying less wasted computational power. Figure~\ref{fig:throughput-breakdown} presents a breakdown study isolating each optimization's contribution. The baseline scale-in policy absorbs a failed node's workload locally within its pipeline stage, creating a persistent straggler that gates throughput and yields low Load Scaling Efficiency (LSE). The most critical optimization, layer migration, resolves this bottleneck by globally redistributing the excess load across data and pipeline parallel dimensions, delivering a dominant improvement in LSE. This global rebalancing is the primary source of performance gain, though its benefits diminish as failures accumulate. Subsequently, DVFS provides fine-tuning by selectively up-clocking residual slow stages, adding another percentage point to the LSE. Combined, these optimizations achieve an LSE of $\ge$ 0.89 at all shrink points, demonstrating near-linear throughput. Layer migration accounts for 80–95\% of the total improvement, with DVFS correcting minor residual imbalances.

\subsection{Overhead of Per-step Snapshot}
\begin{table}[htbp]
\centering
\caption{Throughput with/without per-step snapshot (Samples per Second).}
\label{tab:zero-ckpt-throughput}
\resizebox{1\columnwidth}{!}{%
\begin{tabular}{lrrr}
\toprule
Model & \begin{tabular}[c]{@{}c@{}}No snapshot \end{tabular} & \begin{tabular}[c]{@{}c@{}}With snapshot\end{tabular} & \begin{tabular}[c]{@{}c@{}}Throughput \\ loss (\%)\end{tabular} \\
\midrule
Llama2-7B  & 51.941 & 51.700 & 0.46 \\
Llama2-13B & 32.805 & 32.602 & 0.62 \\
Llama2-34B &  8.545 &  8.487 & 0.69 \\
\bottomrule
\end{tabular}%
}
\end{table}
Per-step snapshotting incurs a negligible performance overhead, with a throughput reduction of less than 1\% across Llama-2 models ranging from 7B to 34B parameters (Table~\ref{tab:zero-ckpt-throughput}). This efficiency stems from a pipelined design that hides snapshot latency within the training's critical path, as shown in Figure~\ref{fig:ckpt_b}. Specifically, gradient transfers (D2D and D2H) are overlapped with concurrent device operations like the local optimizer step and All-Gather, while the final host-side parameter update is concealed by the subsequent training iteration. This ensures the backup process does not stall computation, and the overhead does not amplify with model size, demonstrating stable scaling.

\subsection{MTTR Analysis}
With online elasticity, most MTTR is eliminated and restart-related downtime disappears; the residual cost primarily comes from communicator reconstruction and system scheduling.

\textbf{Communication Initialization.} Current failure recovery relies on rebuilding communication groups—either a \textit{full restart, which rebuilds the global communicator}, or a \textit{partial restart, which rebuilds only the groups involving the failed node}. Both methods incur significant overhead. As shown in Figure~\ref{fig:comm-group-update}, our \textit{Dynamic Communicator}’s localized design, which avoids costly global group reconstruction by only editing communication links adjacent to the failed rank, achieves a sub-second, near-constant recovery time of 0.15–0.37 s across 8–64 ranks. This yields speedup of 38–82$\times$ over a full restart (12–16s) and 2.8–3.6$\times$ over a partial restart (0.54–1.09s).

\begin{figure}[!htbp]
  \centering
  \includegraphics[width=\columnwidth]{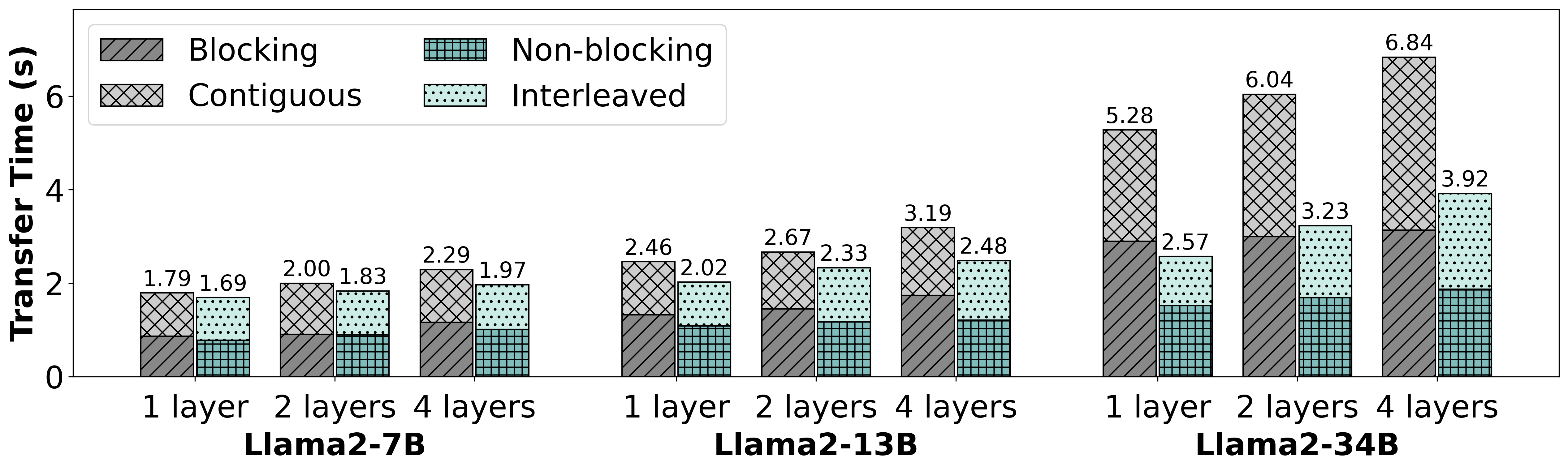}
  \caption{MTTR for layer migration on Llama-2 models, comparing our optimized design (non-blocking parameter migration, interleaved ZeRO layout) against a baseline (blocking copy, contiguous layout) when moving 1, 2, or 4 layers.}
  \label{fig:mttr-layer-transfer}
\end{figure}

\textbf{Parameter Migration.} Our optimizations for layer migration significantly reduce MTTR, with gains that amplify with model size (Figure~\ref{fig:mttr-layer-transfer}). The total MTTR is reduced by 6–14\% for the Llama-2-7B model, 13–22\% for 13B, and a substantial 43–51\% for 34B.

\textit{Model Parameter Migration.} Our non-blocking approach reduces model parameter migration time with benefits that amplify at scale (Figure~\ref{fig:mttr-layer-transfer}). The average MTTR reduction grows from a modest 8\% for Llama-2-7B to 22\% for 13B and 44\% for 34B. This scaling advantage occurs because for larger models, data migration time dominates fixed orchestration costs. Our non-blocking design mitigates this dominant cost by overlapping the data transfer with other critical-path operations, effectively hiding the latency.

\textit{Optimizer Parameter Migration.} For optimizer states, our interleaved ZeRO layout is 1.8–2.3$\times$ faster than the contiguous baseline on the 34B model (Figure~\ref{fig:mttr-layer-transfer}). This advantage comes from converting the migration into parallel, rank-to-rank sends, which avoids the costly re-sharding and data compaction of the contiguous layout. The result is reduced communication volume and fewer network hotspots, making it essential for efficient elasticity at scale.
\begin{table}[b]
\centering
\caption{Downstream task results with and without RNG Resharding. The Reduction is calculated as $1 - |\text{diff}|_{\text{RNG}} / |\text{diff}|_{\text{no-RNG}}$.}
\vspace{-1em}
\label{tab:rng-resharding}
\resizebox{\columnwidth}{!}{
\begin{tabular}{l|c|cc|ccc}
\toprule
Task & Original Perf. & \multicolumn{2}{c}{Perf. w/o RNG Resharding} & \multicolumn{3}{c}{Perf. w/ RNG Resharding} \\
\midrule
& acc & acc & |diff| & acc & |diff| & \textbf{Reduction} \\
MMLU~\cite{hendryckstest2021} & 46.18 & 46.03 & 0.15 & 46.20 & \textbf{0.02} & \textbf{86.7\%} \\
BoolQ~\cite{clark2019boolq} & 78.13 & 78.38 & 0.25 & 78.32 & \textbf{0.19} & \textbf{24.0\%} \\
BBH~\cite{srivastava2022beyond} & 35.90 & 36.20 & 0.30 & 35.70 & \textbf{0.20} & \textbf{33.3\%} \\
AGIEval~\cite{zhong2023agieval} & 23.91 & 24.08 & 0.17 & 24.00 & \textbf{0.09} & \textbf{47.1\%} \\
CEval~\cite{huang2023ceval} & 34.03 & 34.92 & 0.89 & 34.62 & \textbf{0.59} & \textbf{33.7\%} \\
\midrule
\textbf{Average} & - & - & - & - & - & \textbf{45.0\%} \\
\bottomrule
\end{tabular}
}
\end{table}

\subsection{Convergence Consistency}\label{subsec:exp_conv}

In experiments, we evaluate the effectiveness of RNG Resharding on the convergence consistency. To do so, we finetuned a Llama2-7B model with LoRA on the GSM8K dataset, using an 8-NPU setup (TP=1, PP=4, DP=2) that scaled down to 7 NPUs to simulate a failure. We first measured the average loss difference ($\mathbb{E}_{\text{step}}[|Loss_\text{normal}-Loss_\text{elastic}|]$) between this elastic run and a no-failure baseline. Without RNG Resharding, the average loss difference was 0.2\%, which dropped to a mere \textbf{0.045\%} with our method, reducing 78\% of the deviation. This improved training stability translates directly to better downstream task performance (Table~\ref{tab:rng-resharding}). Averaged across all benchmarks, the absolute difference of accuracy deviation from the no-failure baseline was \textbf{45\%} lower with RNG Resharding, confirming its effectiveness in preserving model quality during elastic scaling.

\begin{figure*}[t]
    \centering
    \includegraphics[width=1\linewidth]{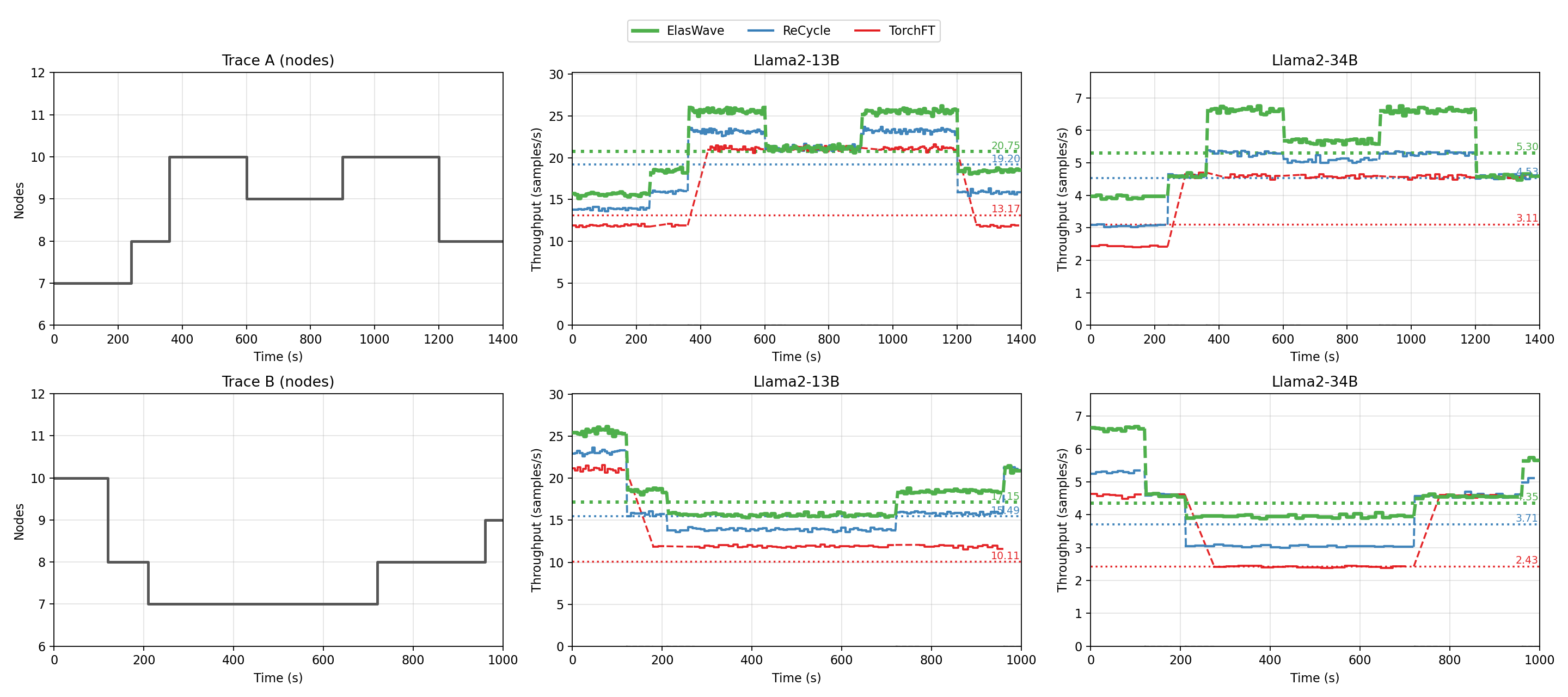}
    \caption{Throughput of \name~(Green), \namerc~(Blue), and \nametft~(Red) on two real-world spot instance traces. Trace A is plateau-heavy and Trace B is shrink-heavy.}
      \label{fig:goodput}
\end{figure*}

\subsection{End-to-End Performance}

On real spot instance traces~\ \cite{miao2024spotserve}, \name~consistently achieves the highest time-averaged throughput, outperforming \namerc~by 10–20\% and \nametft~by 50–70\% across all models and traces. This advantage stems from its coordinated DP+PP rebalancing, which restores near-linear steady states after capacity changes. \nametft’s full restarts lead to a long MTTR ($\approx$ 20s) and the lowest performance. This ranking proves robust across diverse trace patterns.

\subsection{Case Study }

\begin{figure}[!htbp]
    \centering
    \begin{subfigure}{0.48\linewidth}
        \centering
        \includegraphics[width=\linewidth]{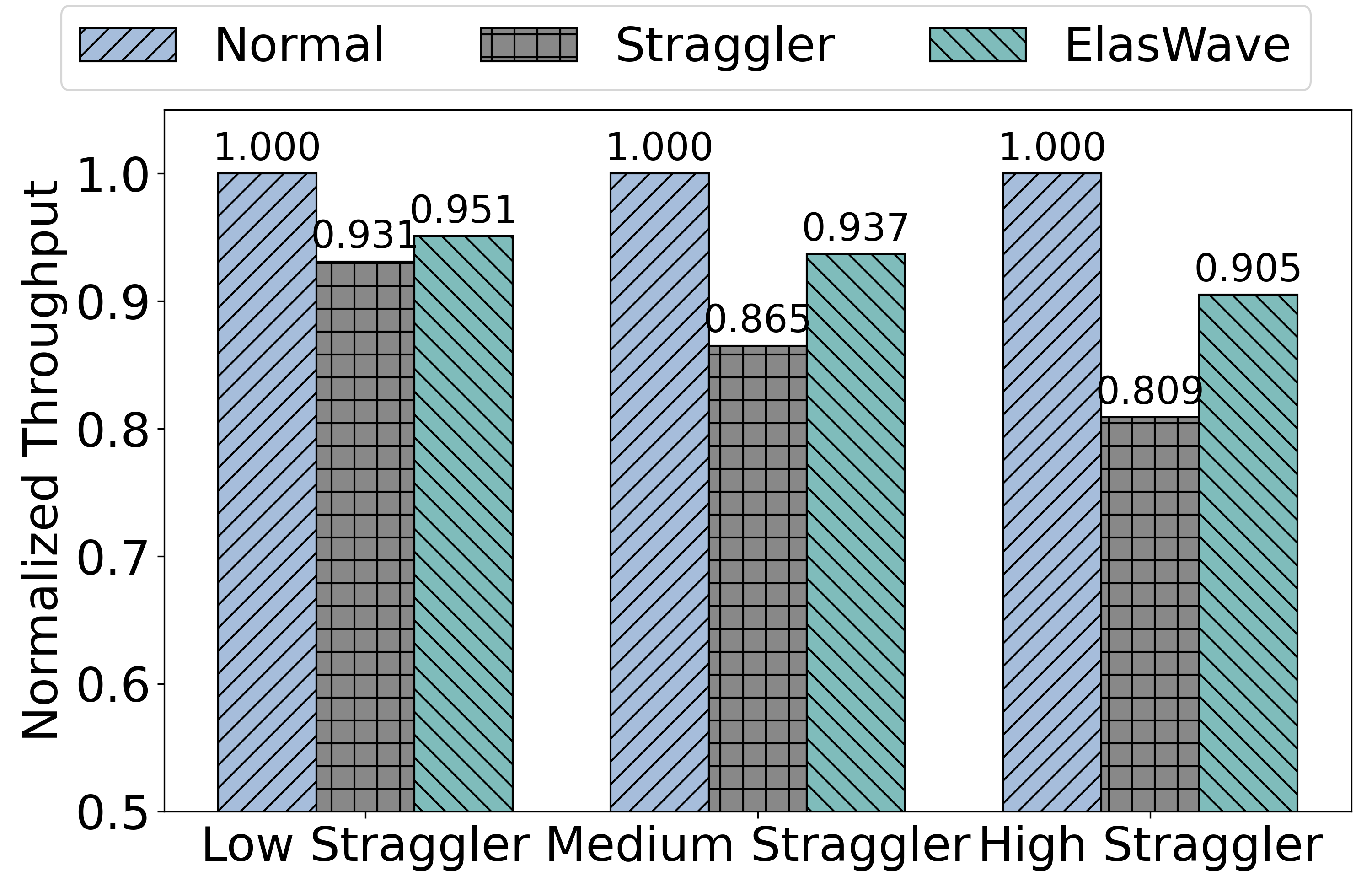}
        \caption{Straggler mitigation under three simulated slowdown levels (Low, Medium, High).}
        \label{fig:straggler-case}
    \end{subfigure}
    \hfill
    \begin{subfigure}{0.48\linewidth}
        \centering
        \includegraphics[width=\linewidth]{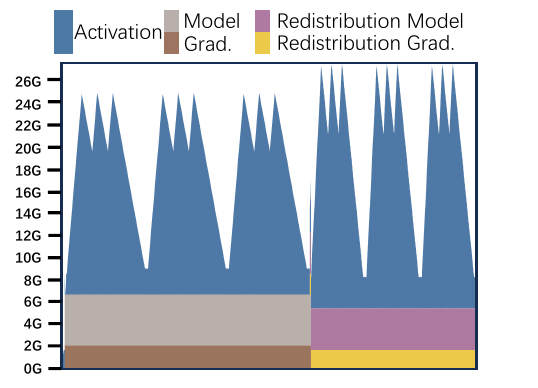}
        \caption{Per-rank memory footprint increases upon cluster shrink due to workload redistribution.}
        \label{fig:memory-change}
    \end{subfigure}
    \caption{\name~performance analysis in case study.}
    \label{fig:case-studies}
\end{figure}

We present two case studies showing \name’s elasticity under both fail-slow conditions and expert-parallel MoE workloads, consistently restoring throughput via rapid rebalancing after perturbations.

\textbf{Fail-slow Mitigation.}
We demonstrate \name's effectiveness in mitigating transient hardware slowdowns (stragglers). We simulate three straggler levels—Low, Medium, and High—by artificially slowing down one worker. As shown in Figure~\ref{fig:straggler-case}, the straggler degrades the normalized throughput to 0.931, 0.865, and 0.809, respectively. By dynamically rebalancing the workload, \name~recovers the throughput to 0.951, 0.937, and 0.905. This corresponds to recouping over 50\% of the performance loss in the medium and high straggler scenarios, showcasing its capability to maintain high efficiency under heterogeneous hardware conditions.

\textbf{Expert Parallelism.} We also evaluate \name~on a Mixture-of-Experts (MoE) model using Llama2-13B, where elasticity is critical for managing expert capacity. In a failure scenario, we compare \name~against a baseline framework, \nametft. After a failure, the baseline's throughput drops to 9.92 samples/sec from an initial 15.73 samples/sec. In contrast, \name's efficient recovery and rebalancing mechanisms achieve a throughput of 13.13 samples/sec, a 32\% improvement over \nametft, recovering a significant portion of the lost performance. This demonstrates \name's superior capability in complex, dynamic workloads like MoE.

\subsection{Discussions}
\name~delivers robust application-level elasticity, handling failures and stragglers by optimizing workload distribution to ensure training continuity. While it masterfully mitigates the \textit{impact} of systemic issues like network congestion, addressing their root cause is beyond its scope. This deliberate focus allows \name~to excel within its application-level domain, providing a resilient core for future co-design with network-aware infrastructure.

\textbf{Extensibility.}The framework's flexibility is further validated on MoE models. That our general-purpose elastic engine functions effectively in such a specialized, dynamic domain is a testament to its robust design. While adapting to SOTA MoE systems like DuoPipe remains future work, our solution provides a strong and versatile foundation for it.

\textbf{OOM Risks.} \name~successfully navigates the fundamental memory trade-offs of elasticity (Figure~\ref{fig:memory-change}). An increased per-rank memory footprint upon cluster shrink is an unavoidable consequence of workload redistribution. \name~adeptly manages these dynamics, absorbing transient activation spikes and guaranteeing against OOM failures. The resulting stable, albeit elevated, memory profile reflects this robust resource management and opens avenues for future work in advanced offloading schemes.

%% file: Contents/9_Related_Work.tex
\section{Related Work}

\stitle{Checkpoint-based fault tolerance.}
\textit{Checkpointing} enables recovery from the most recent stable state by periodically saving the training state. Recent work focuses on mitigating the significant overhead of this process, including asynchronous~\cite{varuna_eurosys22,nicolae2020deepfreeze,wang2024fastpersist,maurya2024datastates}, lightweight in-memory~\cite {wang2023gemini}, incremental checkpoints~\cite{eisenman2022check,agrawal2024inshrinkerator}, and adaptive adjustment of checkpoint frequency~\cite{mohan2021checkfreq,li2025convergence,liu2025checkflow,yao2025lowdiff}. 
To minimize MTTR, \name~skips checkpoint-based rollback and performs online elastic recovery from in-memory step state.

\stitle{Elastic training.}
\name~targets \emph{efficient elastic execution under failures} that natively satisfies multiple metrics and jointly minimizing MTTR and maintaining high post-change throughput while preserving statistical validity.

\textit{MTTR-oriented recovery.} Mainstream toolchains implement elasticity as \emph{restart-and-resume} from checkpoints or in-process state commits (coarse job-level granularity); in-memory checkpointing reduces I/O but still triggers broad pauses and communicator rebuilds, yielding only coarse control over MTTR and post-recovery throughput~\cite{torchelastic,sergeev2018horovod,rasley2020deepspeed,wang2023gemini}. Online reconfiguration for anticipated/detected failures (e.g., spot/preemptible) improves system availability, but MTTR remains coupled to cluster scale and the chosen granularity~\cite{torchft_blog,thorpe2023bamboo,oobleck_sosp23,recycle_sosp24}.

\textit{Throughput-oriented elasticity.} Prior work boosts \emph{steady-state throughput} by reshaping parallelism or injecting redundancy under resource changes or failures. Approaches range from job/stage template switching (pipeline templates) to DP-group rerouting and micro-batch rebalancing; typical granularities are job-level~\cite{athlur2022varuna}, stage/pipeline-level~\cite{thorpe2023bamboo,oobleck_sosp23}, DP-group-level~\cite{torchft_blog}, and micro-batch-level~\cite{recycle_sosp24}. These methods sustain progress but can incur overhead in failure-free periods (redundant compute) and suffer MTTR degradation under heterogeneity or frequent reconfiguration; they also often optimize only a \emph{single} dimension (e.g., DP routing) rather than coordinating data/model/hardware jointly.

\textit{Convergence under elasticity.} Some efforts preserve accuracy consistency during elasticity (e.g., DP resizing rules), while largely assuming \emph{DP-only} elasticity and leaving cross-dimension rebalancing (DP/PP/micro-batch) unresolved~\cite{li2022easyscale,hwang2021elastic,li2023lyra,pollux_osdi21,gu2023elasticflow}.

\stitle{Failure Detection.} 
Effective \textit{fault detection} is essential to minimize downtime and error propagation, leveraging tools like NVIDIA DCGM~\cite{nvidia-dcgm} and mechanisms such as timeouts, heartbeats, and log analysis~\cite{jiang2024megascale,barletta2024mutiny,liu2023predicting,li2023workload,zhang2025llm,guan2024logllm}. In addition, the \name~Agent includes hardware- and process-level detectors on Ascend clusters. Due to page limits, we omit details.

\stitle{Resilience–Convergence Tradeoffs.} 
Several fault-tolerant strategies trade statistical convergence for progress under failures or delays. \textit{Local/periodic-averaging SGD} reduces synchronization to mask stragglers, but client drift and heterogeneity yield slower or biased convergence and accuracy gaps~\cite{haddadpour2019local, karimireddy2020scaffold, li2019convergence}. \textit{Buffered asynchronous aggregation} admits clients as they arrive to sustain throughput, but the resulting asynchrony and client drift degrade statistical efficiency~\cite{nguyen2022federated}. These trade-offs are misaligned with frontier-scale pretraining, where accuracy, stability, and reproducibility are first-class requirements. In contrast, \name{} couples resilience scheduling with RNG-state consistency to maintain optimization fidelity despite failures.


%% file: Contents/10_Conclusion.tex
\section{Conclusion}
We presented \name, an elastic-native LLM training system that performs per-step fault tolerance via multi-dimen\allowbreak sional scheduling across dataflow, graph, DVFS, and RNG. The design couples an in-place dynamic communicator with non-blocking layer migration, interleaved ZeRO state movement, and snapshot-based live remap to meet the four production goals: parameter consistency, low MTTR, high post-change throughput, and computation consistency. The system is currently in a preliminary experimental stage and will undergo further evaluation at extensive scale. On our testbed, \name~improves throughput by up to $1.60\times$ than \nametft~and up to $1.35\times$ than \namerc. Communicator recovery MTTR is improved by up to $82\times$ and $3.6\times$ compared with full and partial rebuilds. Non-blocking migration with interleaved ZeRO cuts Layer migration MTTR  by up to 51\% compared with blocking migration using default ZeRO. RNG Resharding reduces convergence deviation by ~78\%. These results demonstrate that \name~delivers fast, consistent, and scalable elasticity for large-model training and provides a practical path to robust pretraining on fluctuating, hyperscale clusters.